# NANYANG TECHNOLOGICAL UNIVERSITY

**PROGRESS REPORT**
**FOR QUALIFICATION CUM PHD CONFIRMATION EXERCISE**

# TIME-SERIES ANALYSIS ON EDGE-AI HARDWARE FOR HEALTHCARE MONITORING

**HU JINHAI**
**G2003695B**

**SCHOOL OF ELECTRICAL AND ELECTRONIC ENGINEERING**

**JUL 2022**

# Abstract

Time series analysis is an essential tool employed in a wide range of applications to understand the underlying data trends or patterns over a period of time, or even to forecast the future values. It is well suitable for non-stationary data, including many biomedical signals such as electrocardiogram (ECG), electroencephalogram (EEG), which are regularly fluctuating over specific periods.

To monitor biomedical signals continuously, current medical practices require patient to wear a portable recording device for 24 hours or longer period. The recorded data are either stored in the device or uploaded to the cloud for clinician to conduct off-line analysis, which is power-hungry and not efficient. Thus, there are rising needs for online real-time signal classification and forecasting to provide early alarm for the symptoms such as arrythmia or cardiac arrest, so that the necessary medical intervention can be triggered in time to improve the therapeutic outcome.

In this project, a time-domain ECG signal analysis model is developed based on a novel dynamically-biased Long Short-Term Memory (DB-LSTM) neural network. This model can perform both ECG forecasting and classification tasks simultaneously. Better than 98% accuracy and less than 10-3 normalized mean square error in forecasting task have been achieved. As for classification task, it reached fast training convergence with higher than 97% accuracy and lower training parameters contrasted to multiple neural network structures. Optimized for hardware implementation, all the network weights are truncated to INT4/INT3 length with compromising the classification training and inference accuracy for 2%/6% respectively, while no accuracy reduction in forecasting application. Comprehensive simulations with multiple ECG datasets had proven the robustness of the proposed model.

The next step is to implement this algorithm into FPGA and CMOS integrated circuit for practical deployment continuous cardiac monitoring. Also, by developing a platform for AI algorithm applying on digital hardware, multiple types of neural networks with selectable weight quantization can be simulated conveniently. Finally, an online training implemented on chips will be realized to perform health care monitoring.

# Acknowledgements

I would like to express my greatest appreciation and sincerest gratitude to my supervisor, Associate Professor Goh Wang Ling, and co-supervisor Dr. Gao Yuan, for their kind support and professional supervision over this project.

I would also like to extend my sincere gratitude to Institute of Microelectronics (IME), A*STAR and NTU VIRTUS for their kind help and technical support.

Finally, my special appreciation to A*STAR AGS for providing me scholarship and opportunities in embarking on my research journey as a Ph.D. student.

## Authorship Attribution Statement

This Qualifying Examination report includes content where I received support from my supervisor(s) and collaborators as follows:

- Associate Professor Goh Wang Ling and Dr. Gao Yuan provided the initial project direction and edited the QE report.
- I prepared the QE report drafts. Revisions and advice were also taken from Dr. Gao Yuan and Associate Professor Goh Wang Ling

# Table of Contents

## Table Captions

# Figure Captions

# Chapter 1

# Introduction

## 1.1 Motivation

Artificial intelligence internet of things (AIoT) is an emerging research area that combines sensor and artificial intelligence to enable real-time intelligent data analysis on edge devices. Many types of data in our daily life are in time series format. For example, the voice signal and the human vital signs such as electrocardiogram (ECG), electroencephalogram (EEG), etc. Time series data have a natural temporal property. Hence, the ability to extract distinct signal features from the time series becomes an interesting topic and it has attracted great research efforts recently.

Cardiovascular diseases (CVDs) such as heart failure and atrial fibrillation are the leading causes of death globally [1]. Electrocardiogram (ECG) is the golden standard to monitor patient's cardiac condition, by measuring the strength and timing of the electrical activity in the heart. by measuring the strength and timing of the electrical activity in the heart [2]. Some of the cardiovascular events such as heart rate variability (HRV) and atrial fibrillation are rare and not predictable. Therefore, long-term continuous ECG monitoring is required to capture such events for cardiac risk evaluation. Current medical practices require patient to wear a portable recording device for 24 hours or longer period. The recorded data are either stored in the device or uploaded to the cloud for clinician to conduct off-line analysis. However, there are rising needs for online real-time ECG forecasting and classification. The purpose is to forecast and detect the onset of symptoms such as arrythmia or cardiac arrest timely, so that early alarm and necessary medical intervention can be triggered in time to improve the therapeutic outcome.

ECG is time-variant signal in nature. To forecast the values of future time steps and classify a time-variant sequence, the algorithm needs to learn from the past data and map them to an output sequence and respective pattern categories. There are a few key ECG anomaly patterns illustrated in Fig. 1.1, including left/right bundle branch block (L/R), atrial premature (A) and premature ventricular contraction (V). Together with the normal mode (N), five categories are required for meaningful ECG forecasting and classification.

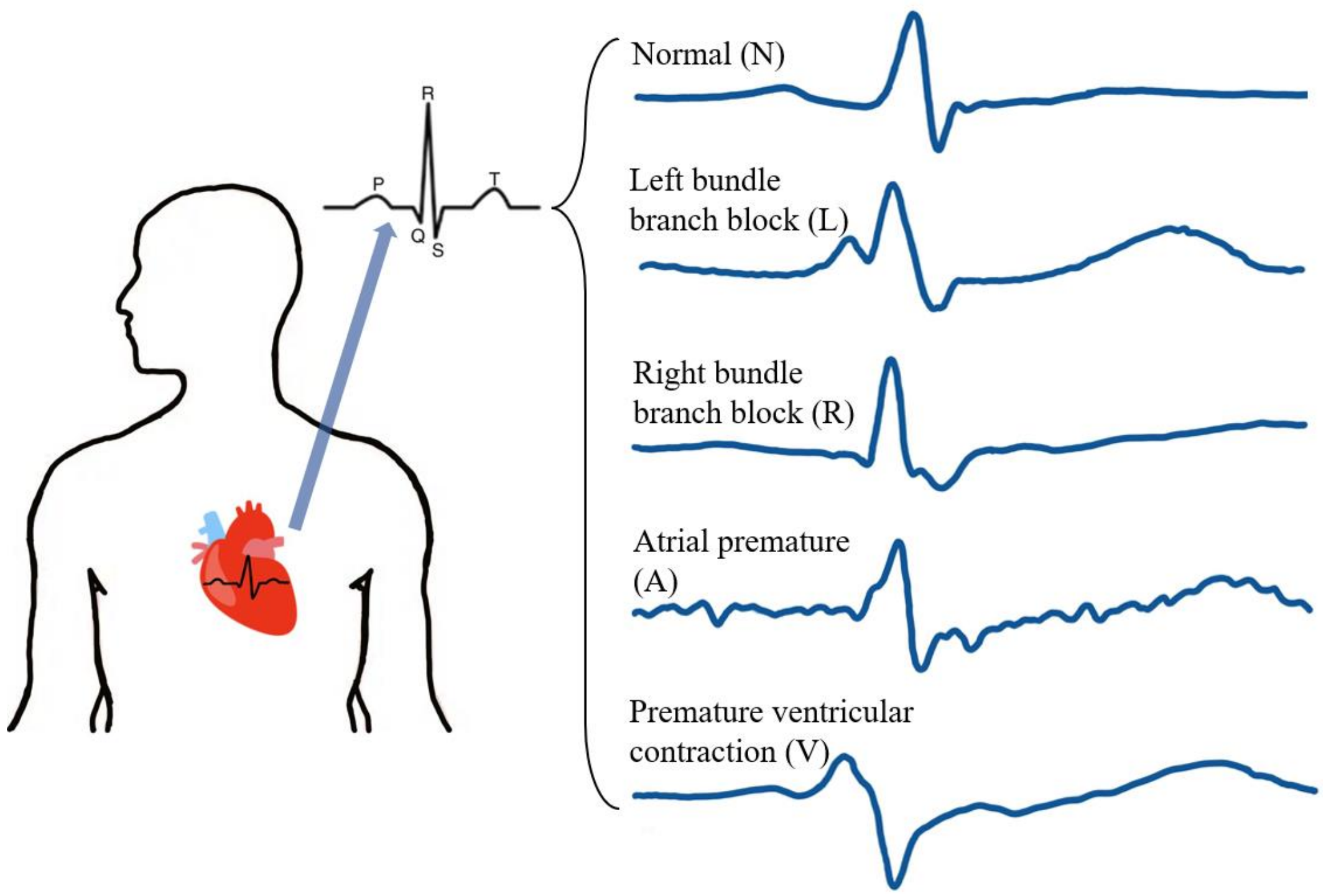

**Figure 1.1:** ECG signal with five key types of waveform pattern.

## 1.2 Objectives

An algorithm level improvement needs to be done to improve the forecasting time steps and accuracy with fixed-point weight precision, as well as to enhance classification accuracy with shorter window size and less training parameters. Thereafter, AI algorithm on hardware should support multiple neural networks for different aspects of application. Following by implementing the AI algorithm on FPGAs or CMOS circuits for health care monitoring. The detailed objectives listed in order:

- Design a model to improve the conventional LSTM algorithm reported as dynamically-biased LSTM (DB-LSTM) [3].
- Implement the DB-LSTM model on ECG signal forecasting and classification applications with MIT-BIH ECG datasets [4].
- Develop DigiNet which a simulation platform for neural networks on digital hardware.
- Integrate the AI algorithms on FPGA for testing and CMOS for tapeout.
- Tapeout a chip that supports online training for healthcare monitoring.

## 1.3 Organization

This report is organized as follows:

- Chapter 2: A literature review for recurrent algorithms and their applications on Bio-signal.
- Chapter 3: Introduces the proposed model and training procedure, as well as fixed-point weight truncation.
- Chapter 4: Perform the results for ECG forecasting and classification.
- Chapter 5: Compares this model to other similar inventions.
- Chapter 6: Make conclusion and explain future works.

# Chapter 2

# Literature Review

## 2.1 Introduction

This chapter consists of two parts of literature review, Recurrent algorithms and ECG applications. Two widely used recurrent algorithms are reviewed and their applications on edge devices are also discussed. On the other hand, many state-of-the-art methods on forecasting and classifying ECG signals are discussed here, where forecasting ability as well as classification latency and accuracy are the evaluation criteria. Moreover, to embed ECG forecasting and classification tasks on edge devices, weight quantization is necessary and important. Thus, the algorithms should be modified and fit into the limit hardware resources of edge device with acceptable accuracy degradations.

## 2.2 Literature Review for Recurrent Algorithms

### 2.2.1 Long Short-Term Memory Algorithm

The Long Short-Term Memory (LSTM) illustrated in Fig 2.1 is a widely used approach in Recurrent Neural Network (RNN). It can solve the problems of gradient disappearance and explosion during long sequence training.

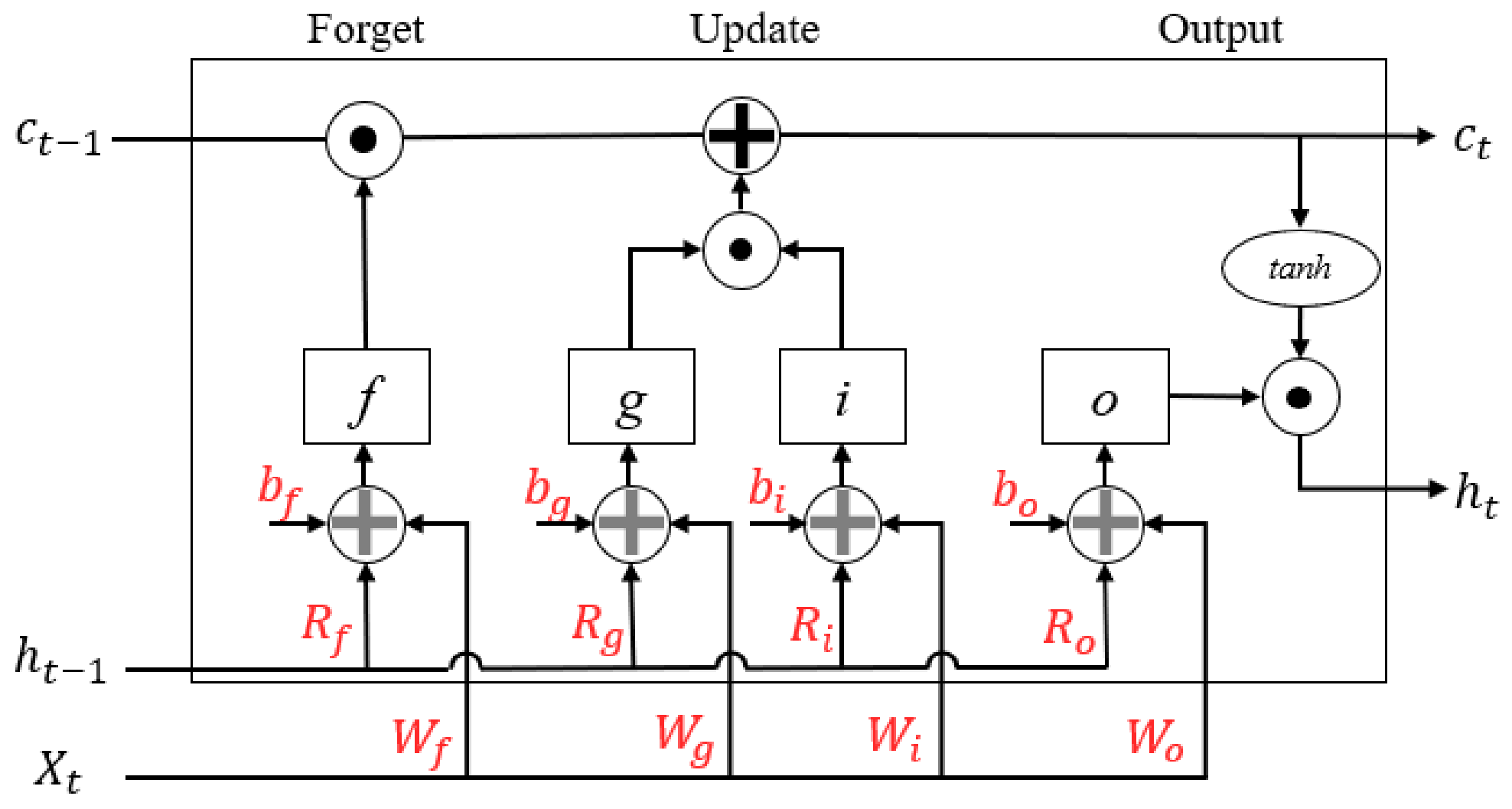

**Figure 2.1:** Conventional LSTM cell structure.

TABLE I LSTM CELL STRUCTURE PARAMETERS

| | |
|---|---|
| $X_t$ | Current input vector |
| $c_t$ | Current cell state |
| $h_t$ | Current cell output (hidden state) |
| $c_{t-1}$ | Previous cell state |
| $h_{t-1}$ | Previous cell output (hidden state) |
| $f_t, g_t, i_t, o_t$ | Current Forget, Generate, Input, Output statues |
| $W_{f,g,i,o}$ | Input weights for *f, g, i, o* |
| $R_{f,g,i,o}$ | Recurrent weights for *f, g, i, o* |
| $b_{f,g,i,o}$ | Bias for *f, g, i, o* |

[5] demonstrated backward pass training to counteract linearly growth in forward pass to allow one dimensional (1-D) long sequence data to fit in the LSTM. Thereafter, a hierarchical multi-dimensional RNN is innovated which extended from unsegmented sequence data labelling [6].

The four statuses of conventional LSTM can be expressed by (1) – (4):

$$f_t = \sigma\left(X_t W_f^T + h_{t-1} R_f^T + b_f\right) \tag{1}$$
$$g_t = tanh\left(X_t W_g^T + h_{t-1} R_g^T + b_g\right) \tag{2}$$
$$i_t = \sigma(X_t W_i^T + h_{t-1} R_i^T + b_i) \tag{3}$$
$$o_t = \sigma(X_t W_o^T + h_{t-1} R_o^T + b_o) \tag{4}$$

where $\sigma$ and *tanh* stands for sigmoid and hyperbolic tangent active function given in (5) and (6), respectively.

$$\sigma(x) = \frac{1}{1 + e^{(-x)}} \tag{5}$$
$$\tanh(x) = \frac{e^{2x} - 1}{e^{2x} + 1} \tag{6}$$

After which, the updating of cell state and selection of hidden state is shown in (7) – (8):

$$c_t = f_t \odot c_{t-1} + i_t \odot g_t \tag{7}$$
$$h_t = o_t \odot \tanh c_t \tag{8}$$

### 2.2.2 Gated Recurrent Unit Algorithm

Recently, [7] utilized multi-dimensional LSTM and achieved good results in multivariate feature applications. However, more researchers are seeking to modify LSTM to enhance its ability in handling complicated sequence data. Thus, Gated Recurrent Unit (GRU) shown in Fig 2.2 and empirical evaluation were established and reported in [8]. Both GRU and LSTM were proven to adapt in different multivariate time series problems based on their performance in [9, 10].

GRU is a compressed model based on LSTM, and its workflow is shown in (9) – (12)

$$z_t = \sigma(W_z \times [h_{t-1}, X_t]) \tag{9}$$
$$r_t = \sigma(W_r \times [h_{t-1}, X_t]) \tag{10}$$
$$g_t = \tanh\left(W_g \times [r_t \odot h_{t-1}, X_t]\right) \tag{11}$$
$$h_t = (1 - z_t) \odot h_{t-1} + z_t \odot g_t \tag{12}$$

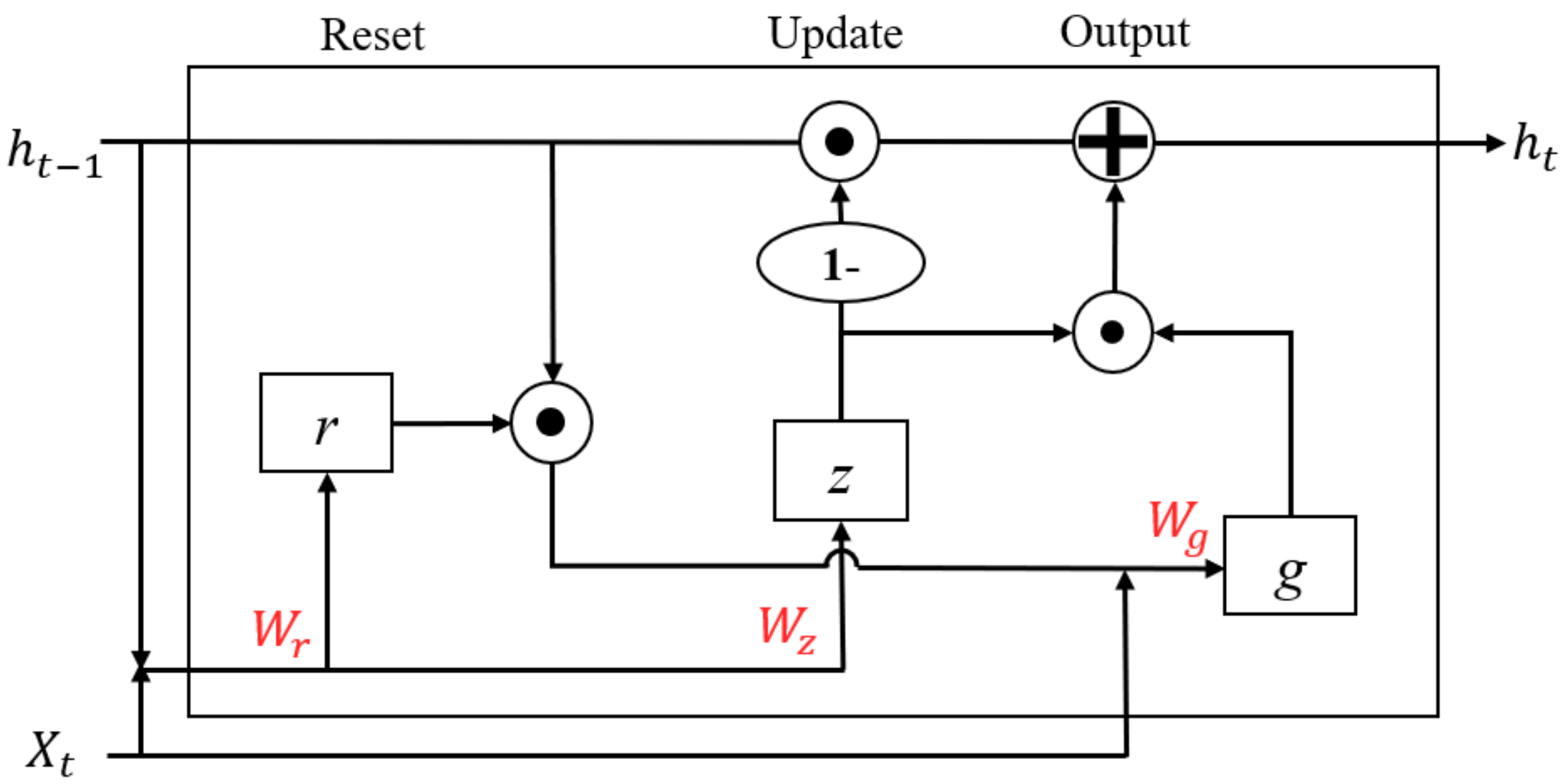

**Figure 2.2:** Conventional GRU cell structure.

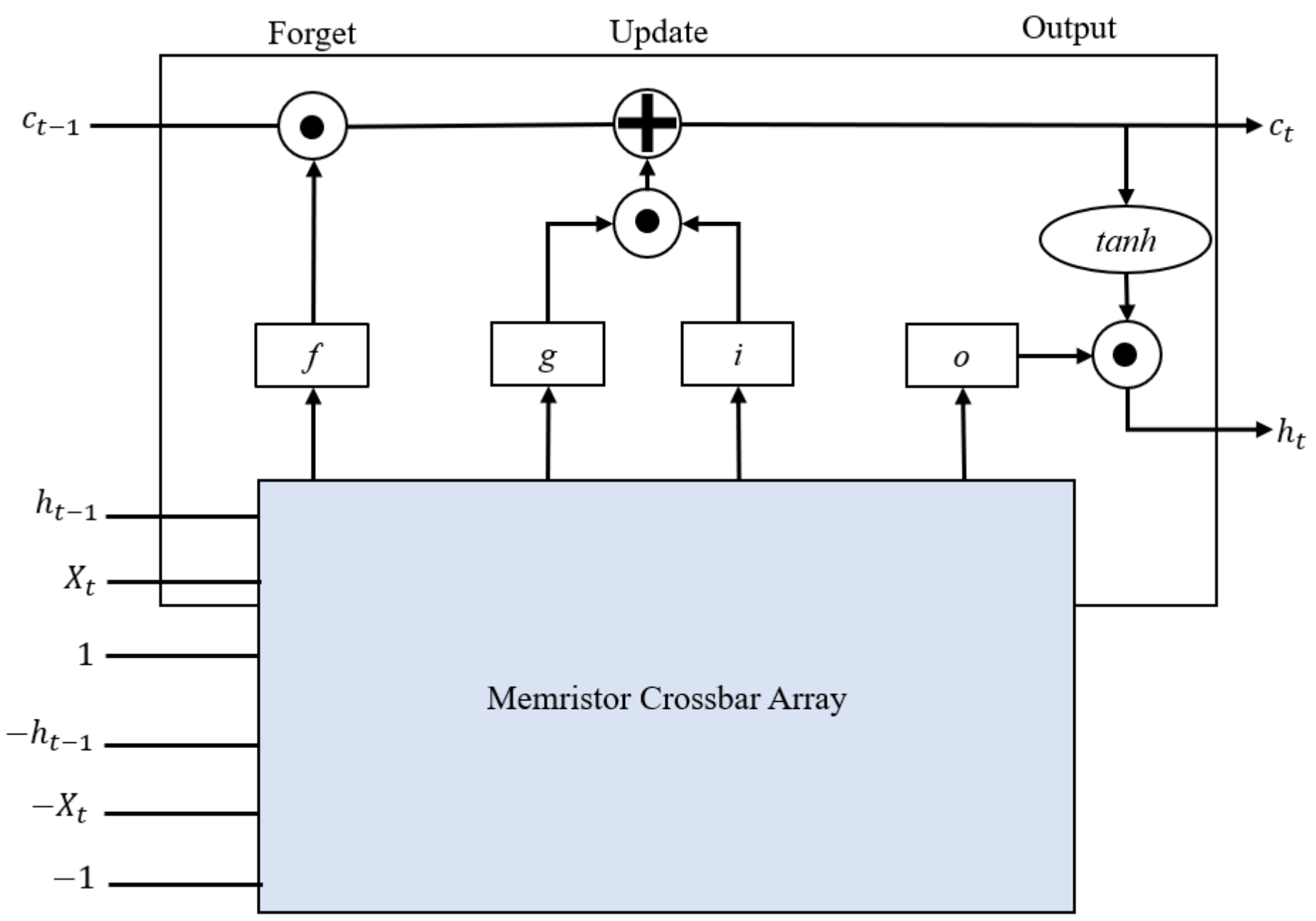

**Figure 2.3:** Memristor crossbar array on LSTM algorithm.

### 2.2.3 Recurrent Algorithm on Edge Devices

On the other hand, edge device has limited hardware resources including computation capacity, memory size and battery lifetime. It remains a great challenge to implement the artificial intelligence algorithm on edge devices. LSTM has been implemented in hardware platforms such as FPGA [11, 12] and memristor crossbar arrays [13] shown in Fig 2.3. Hence, it is vital and necessary to design a novel cell structure for LSTM with both floating and fixed-point weight design, that can enhance the neural network performance on software level and also be applied and operated on edge devices.

## 2.3 Literature Review for Bio-signal Applications

There are many prior works on applying various machine learning algorithms to analyse ECG data for cardiac risk predication and classification [14-24].

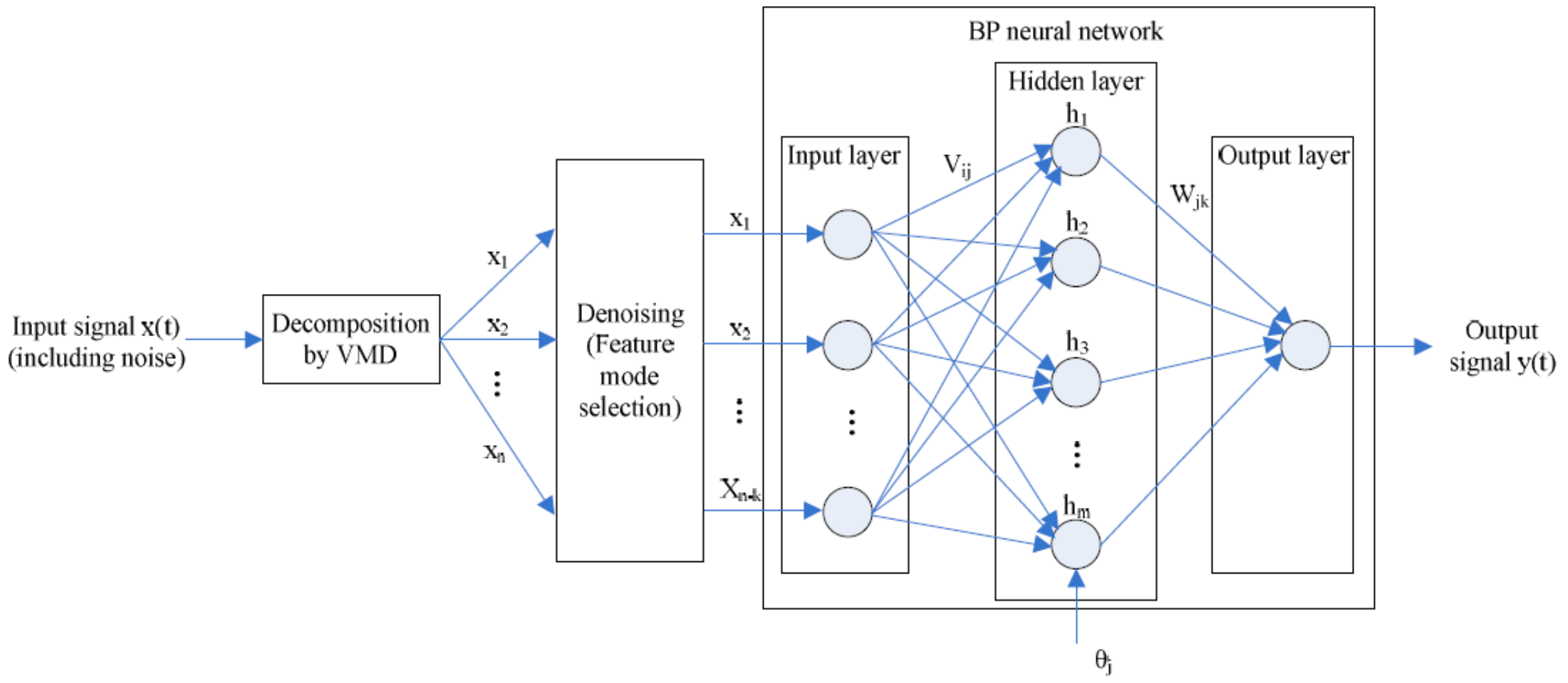

**Figure 2.4:** ECG prediction model with VMD+NN. [Taken from Z. Sun, Y. Lei, J. Wang, Q. Liu and Q. Tan, ICEIEC 2017]

### 2.3.1 Forecasting of Bio-signal

The recurrent neural network can by applied on many 11-dimentional time-series forecasting on bio signals, such as the ECG, Electroencephalogram (EEG) and surface electromyography (sEMG). Signal processing involving Fourier transform (FT) combined with multi-layer perceptron neural network [16, 17] provided a hybrid system for ECG forecasting. A TS fuzzy control system involving Gaussian membership function to do single time step forecasting is illustrated in [18]. Although they all yielded low Root Mean Square Error (RMSE), multiple FT process and Gaussian operations enlarge the time complexity and only one time step forecasting can be realized. Variational Mode Decomposition combined with Neural Network (VMD+NN) model [16], shown in Fig 2.4, provides a hybrid system for ECG prediction, from which VMD is to decompose the ECG original signal into nine sub signals based on frequency domain and then send to the three-layer BPNN consisting of 180 learnable parameters. Nine segments of 2160 samples are trained and resulted 0.0233 RMSE and 98.8% accuracy. A similar model [17] that combined phase space reconstruction and NN including 14 input neurons, 20 hidden neurons and single output neuron is applied to the same dataset, which performed 0.0423 RMSE and 97.3% accuracy. [18] illustrated a non-neuron method to do single time step forecasting for ECG signal. A fuzzy control system with 3600 samples is used to obtain the TS fuzzy model involving Gaussian membership function. This model gave a high performance of 0.0146 RMSE, but multiple Gaussian operations enlarge the time complexity and only one time step forecasting can be realized.

### 2.3.2 Classification of ECG Signal

Convolutional neural network (CNN), both in one dimensional (1D) [19-23] and two-dimensional (2D) [24] formats have been applied in ECG classification analysis. [19] and [20] identified the position of peaks within ECG signal, which built the foundation of heartbeat classification. [22] proposed a hierarchy system which first classify heartbeats into normal or abnormal, and further classify abnormal into four different categories, whose TNSS structure is shown in Fig 2.5. The complex system achieved satisfied accuracy and energy efficiency at a certain level, but the drawback is that it required huge number of weight parameters, which resulted in large memory size if implemented on hardware devices.Pure 1D-CNN consumed less training parameters [21, 23], but required more layers and longer window size to trade off the classification results. Fig 2.6 shows the model architecture as well as its training performance of [23]. Its training tendency converged fast within 100 epochs but not stable eventually, which would cause uncertainty for applying this algorithm on edge devices. In 2D-CNN, ECG signal is converted to image and then perform image processing [24, 25]. To achieve satisfactory results, large-scale deep CNN is required. For example, the CNN model in [24] contains more than 10 million weight parameters and requires huge computation power, therefore it is very challenging to implement such kind of algorithm in resource constraint portable devices. On the other hand, Long Short-Term Memory (LSTM) [26], which is a kind of recurrent neural network, has demonstrated superior forecasting and classification accuracy in time series analysis. Although LSTM can achieve similar performance with smaller network size compared to CNN, deeper networks such as bidirectional LSTM are required for practical ECG applications [27], which is still a huge computation burden for edge devices.

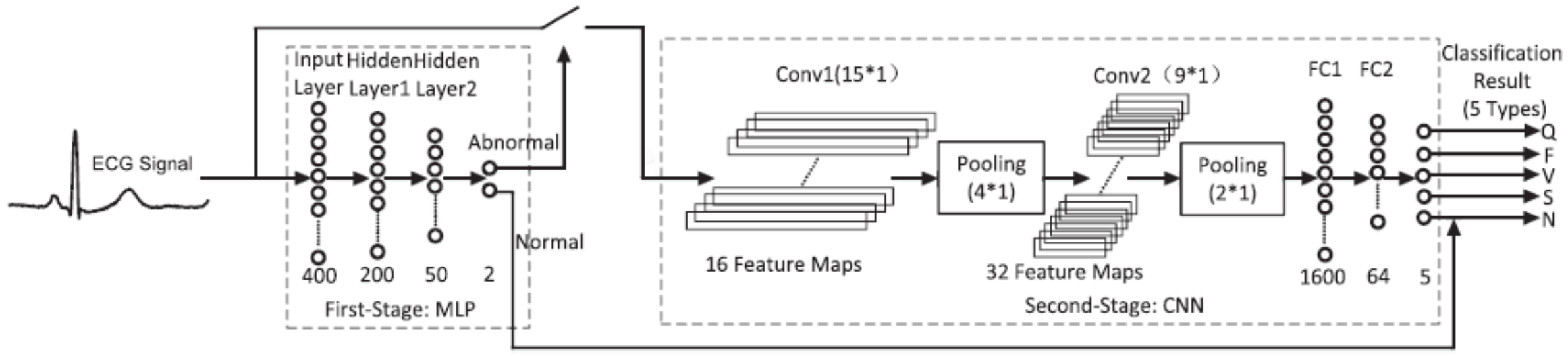

**Figure 2.5:** TSNN structure for ECG classification. [Taken from N. Wang, J. Zhou, G. Dai, J. Huang and Y. Xie, TBioCAS 2019]

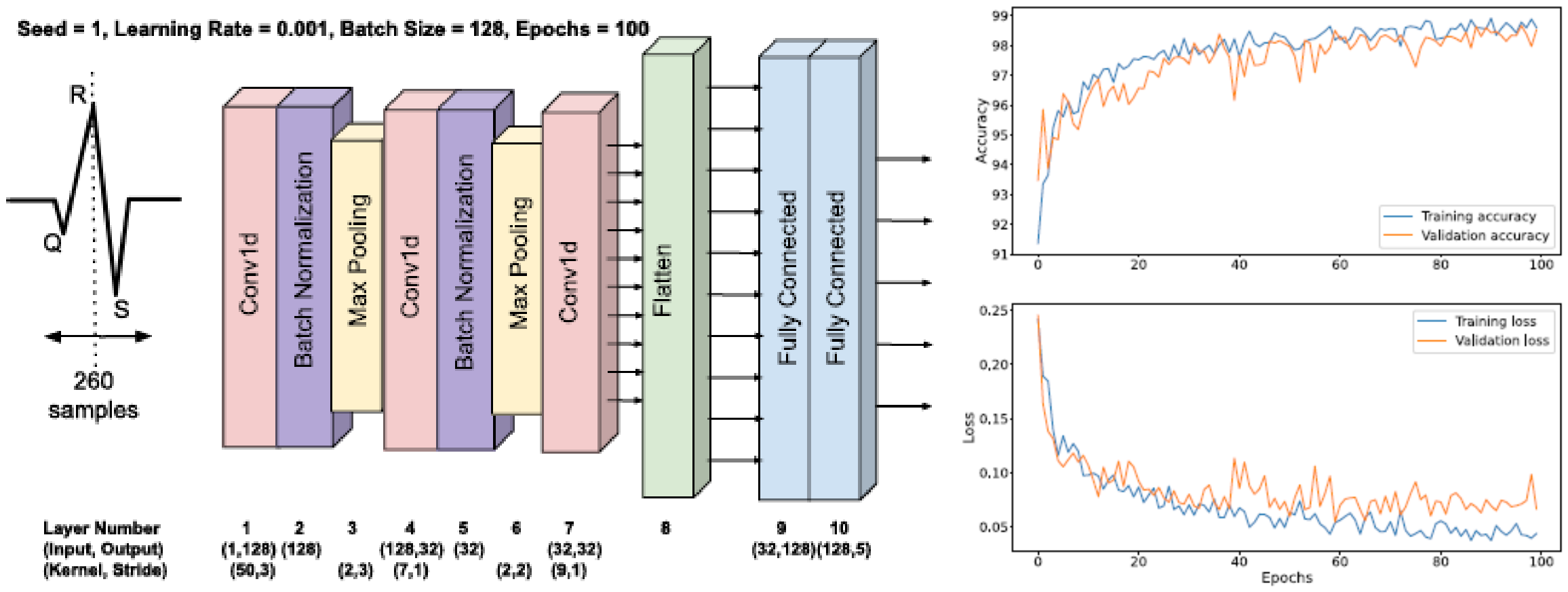

**Figure 2.6:** 1-D CNN for ECG classification and its performance. [Taken from X. Li, et al, ICECS 2021]

In addition, the conventional LSTM and CNN performance will degrade significantly when the weights are truncated to lower resolution for hardware implementation. Longer or even nonconvergence of training processes may occur, resulting in stability issue when on-line training is necessary to build personalized ECG prediction and classification model.

# Chapter 3

## Methodology

The proposed dynamically biased LSTM cell structure is illustrated in Fig. 3.1, and its corresponding notations of parameters are listed in Table II. In the following sub sections, the feedforward propagation, backpropagation and fix-point weight training for both forecasting and classification applications are discussed to describe the entire properties of DB-LSTM cell structure.

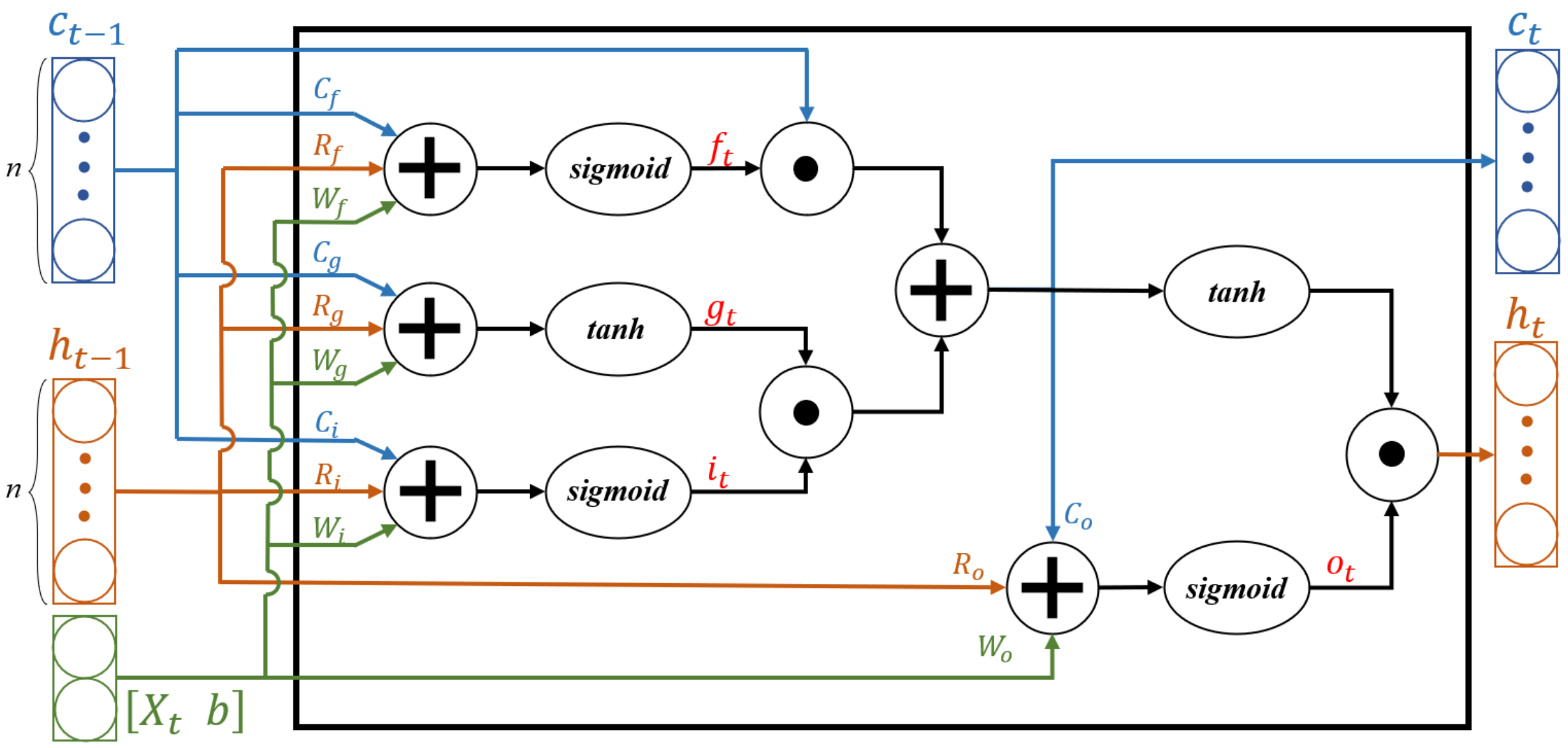

**Figure 3.1:** Dynamically biased LSTM cell structure.

TABLE II DB-LSTM CELL STRUCTURE PARAMETERS

| | |
|---|---|
| $X_t$ | Current input vector |
| $b$ | Bias |
| $c_t$ | Current cell state |
| $h_t$ | Current cell output (hidden state) |
| $c_{t-1}$ | Previous cell state |
| $h_{t-1}$ | Previous cell output (hidden state) |
| $h_{end}$ | Cell ouput (hidden state) at last time step |
| $f_t, g_t, i_t, o_t$ | Current Forget, Generate, Input, Output statues |
| $W_{f,g,i,o}$ | Input weights for *f*, *g*, *i*, *o* |
| $R_{f,g,i,o}$ | Recurrent weights for *f*, *g*, *i*, *o* |
| $C_{f,g,i,o}$ | Cell weights for *f*, *g*, *i*, *o* |
| $W_{oh}$ | Weights for fully connected layer |
| $Out$ | Output value |
| $Prob$ | Probability of each class in classification model |

## 3.1 Feedforward Propagation (FP)

### 3.1.1 FP for Forecasting Model

The feedforward propagation of forecasting model is a three-layer neural network shown in the upper branch of Fig. 3.2, including a pre-processing and input layer which normalize and denoise the ECG signals, a DB-LSTM layer to perform forecasting for next heartbeat cycle and send to the output layer. Through 0.04 threshold for filtering of Discrete Wavelet Transform, the original ECG wave is denoised and smoothed, which can reduce the oscillation effect to the learning system. Since most types of ECG signal shown in Fig. 1.1 consist of extreme R peak value, z-score normalization can maintain useful information about outliers and make the algorithm less sensitive to them in contrast to min-max scaling.

The input layer receives the pre-processed ECG signal with m features and window size of k to be trained. It embeds a shareable random number as dynamic bias, which is split into four different biases due to the $W_{f,g,i,o}$ adjustment throughout the training. Thus, the bias can be fine-tuned, providing two-fold advantages, for the gate output to be more accurate, and also to smooth the back-propagation loss gradient. Thereafter, the processed input vector is sent to the DB-LSTM layer that is made up of single DB-LSTM cell consisting of four statuses, that discards non useful message, update memory, and control the output, respectively. At time step t, the input vector is $[X_t \;\; b]$ with dimension $(m+1) \times 1$, and its corresponding $W_{f,g,i,o}$ has the size $n \times (m+1)$. If the sequence length is set to k, the total training steps within one epoch is k. When the current time step t is less than k, the output of the DB-LSTM cell will be redirected to the recurrent input in the next time step. Otherwise, the hidden state $h_t$ is sent to the output layer, whose dimension is $n \times 1$. As a result, the input and output matrices of the neural network are $(m+1) \times k$ and $n \times k$, respectively.

The inner feedforward processing of DB-LSTM cell can be expressed as (13) – (16). The cell state and $C_{f,g,i,o}$ are the showcases for advanced processing in this model. Different from the conventional LSTM structure, *f*, *g*, *i*, *o* statuses have one additional parameter, that is, the cell state. *f*, *g* and *i* gates consider both $h_{t-1}$ and $c_{t-1}$, while *o* gate counts in the current cell state $c_t$ instead of $c_{t-1}$.

$$f_t = \sigma\left(W_f \times [X_t \;\; b] + R_f \times h_{t-1} + C_f \times c_{t-1}\right) \quad (13)$$

$$g_t = \tanh\left(W_g \times [X_t \;\; b] + R_g \times h_{t-1} + C_g \times c_{t-1}\right) \quad (14)$$

$$i_t = \sigma(W_i \times [X_t \;\; b] + R_i \times h_{t-1} + C_i \times c_{t-1}) \quad (15)$$

$$o_t = \sigma(W_o \times [X_t \;\; b] + R_o \times h_{t-1} + C_o \times c_t) \quad (16)$$

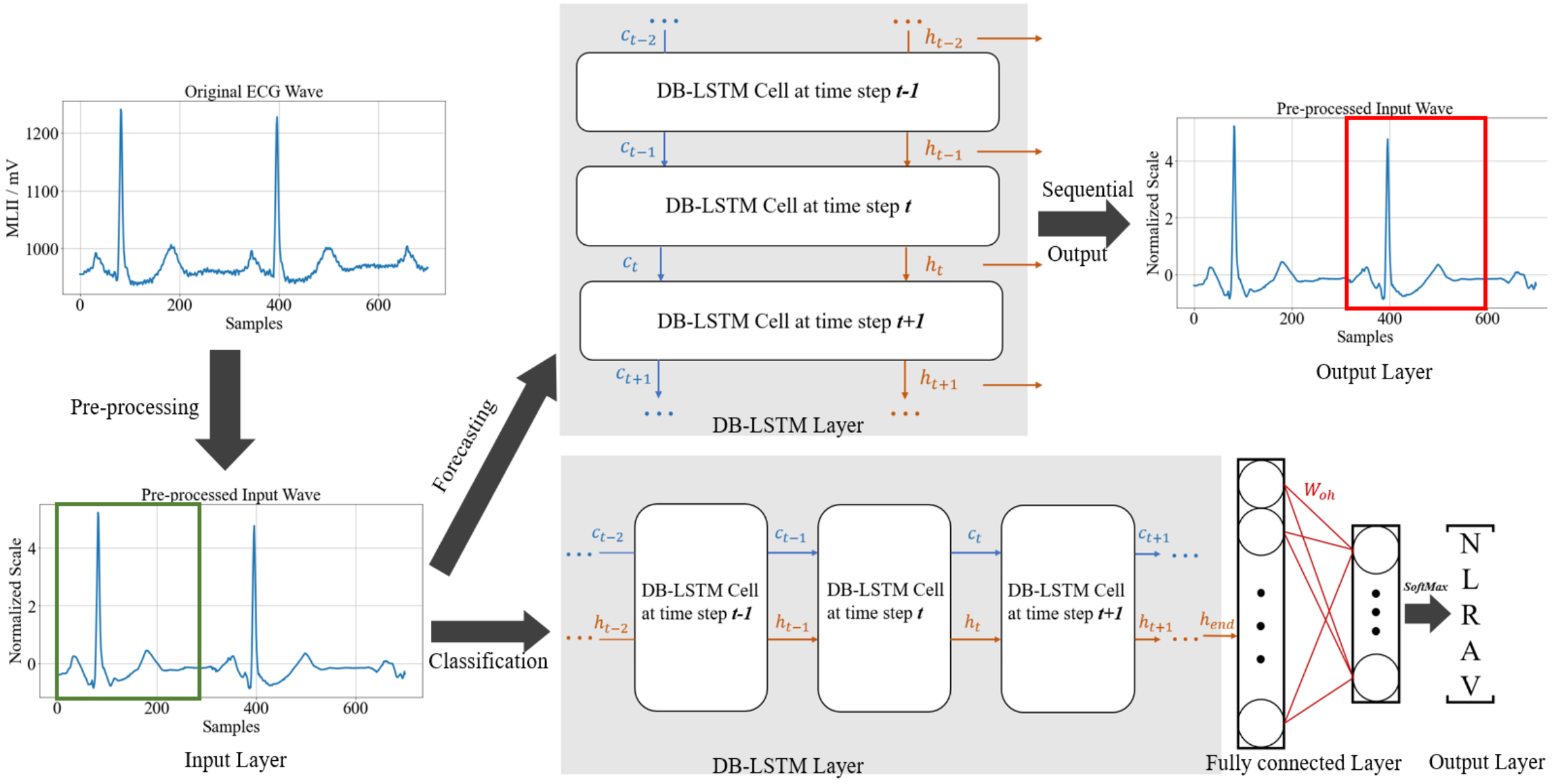

**Figure 3.2:** Block diagram of the proposed ECG monitoring algorithm architecture.

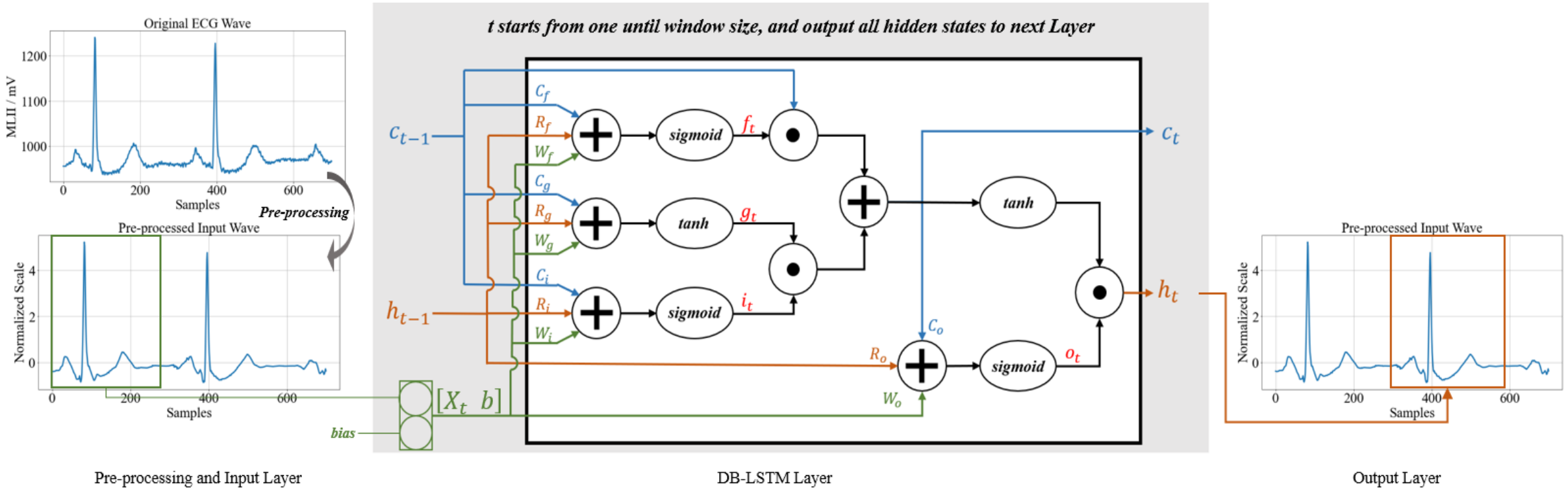

**Figure 3.3:** Block diagram of the proposed DB-LSTM forecasting algorithm model.

where σ is the sigmoid function, and × stands for vector matrix multiplication.

Since the hidden state $h_t$ is a selected and scaled result from the cell state $c_t$, it may cause a huge difference upon final decision. As a result, $f$ and $i$ gates will analyse current input, previous cell state and the output to yield a moderated identification on the level of old messages and the latest information that the LSTM cell should discard or remember in. Similarly, the output gate $o$ also monitors the current cell state to determine what needs to be sent to the hidden output. The difference between the selection of $c_t$ and $c_{t-1}$ is that o is the gating machine working on the process from the current cell state to the hidden state, while $f$ and $i$ are deciding on how the old data can affect the current $c$. The Generation Status $g$ prepares a group of preliminary information that is to be added into the cell state, thus, tanh is selected as the active function. According to the LSTM flow chart shown in Fig. 3.1, the cell state is updated firstly by the forget status followed by the input status.

Also, a single directional LSTM can only gain the learnable features from its previous information and cannot be influenced by the later key words. This is resolved by using bi-directional LSTM [27]. However, bi-directional LSTM doubles the size of all types of weight in its cell structure, which will require more memory space in hardware implementation. Therefore, in this model, to prevent the forget gate to accidentally abandon some useful data that possibly can have a strong connection with the later information, the add-on $c_{t-1}$ in g will cause the update process to re-consider some data stores in the previous cell state, so that it can recover those data and decline initial information loss during the first epoch training. Moreover, the back-up data sent to $g$ is filtered by its weight $C_g$, and the recovered message is determined by the input data and gating signal $i$. The update of cell state is given by (17)

$$c_t = f_t \odot c_{t-1} + g_t \odot i_t \tag{17}$$

where ⊙ is the Hadamard product that states every entry in the old cell state and needs to be updated to attain the new cell state. (18) is the last step in feed forward pass to compute the hidden output.

$$h_t = \tanh(c_t) \odot o_t \tag{18}$$

In summary, the input parameters decrease due to the injection dynamic bias, which provides a better adjustment to the statuses' operations. The cell state enables the gating selection to be more accurate and also resilient to random events. And the cell state parameter in $g_t$ allows LSTM to hold a larger memory and reduce the effect of single direction LSTM's deficiency. In forward pass, (13) – (15) are parallel operations, while (16) – (18) are series operations. Thus, the current process gains the memory with more than one step as compared to the previous since $c_{t-1}$ is computed in advance of $h_{t-1}$. The detailed schematic diagram can be found in Fig 3.3.

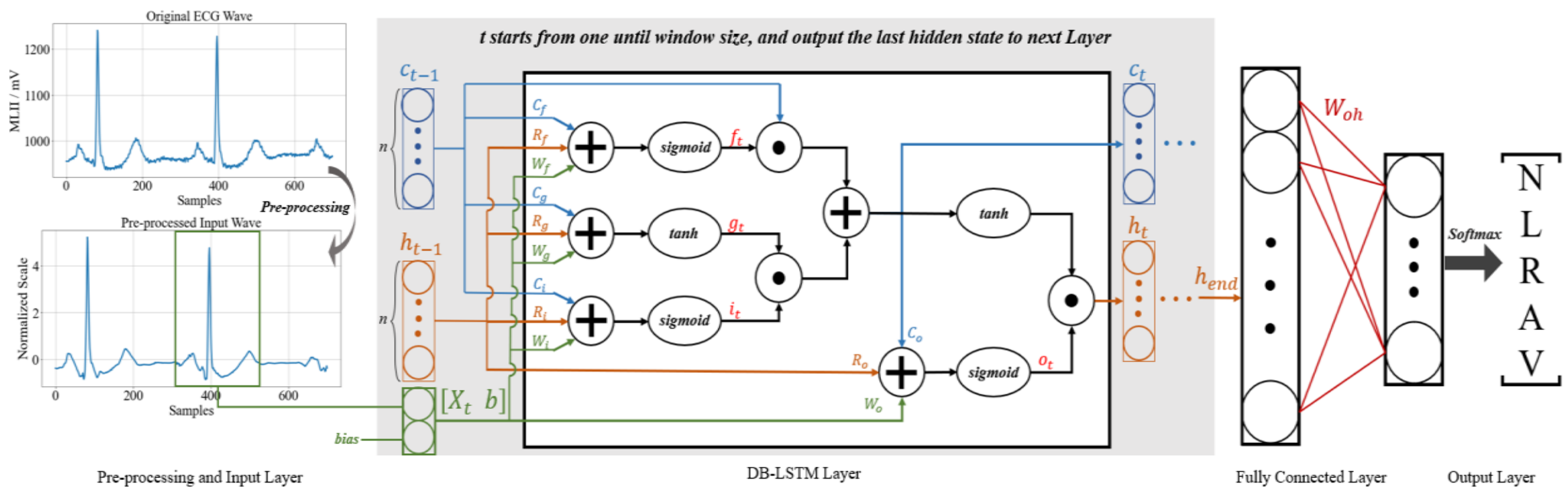

**Figure 3.4:** Block diagram of the proposed DB-LSTM classification algorithm model.

### 3.1.2 FP for Classification Model

The difference between forecasting and classification model is that classification model embeds a fully connected layer after DB-LSTM layer to classify the hidden features into target classes shown in the lower branch of Fig. 3.2, whose equation is shown in (19). For multi-class classification task, more than single hidden neuron need to be used to ensure high qualified feature extraction, and the influence on accuracy based on various hidden number will be discussed in Chapter 4. Moreover, when the current time step t is less than k, the output of the DB-LSTM cell, $c_t$ and $h_t$ with dimension of $n \times 1$, will be redirected to the recurrent input in the next time step, as shown in the shaded block in Fig.3.4. Otherwise, the final hidden state $h_{end}$ is sent to the fully connected layer with n neurons to concentrate these n hidden neurons into five output neurons, whose weight $W_{Oh}$ has size $5 \times n$. The output layer will perform SoftMax function to the five output neurons and generate five categories. As a result, the input and output matrices of the neural network are $(m+1) \times k$ and $5 \times 1$, respectively.

$$Out = W_{Oh} \times h_{end} \tag{19}$$

The final probability of a piece of ECG signal into five classes can be obtained by send the output of FC layer through a SoftMax function as shown in (20)

$$Prob = \frac{e^{Out}}{\sum e^{Out}} \tag{20}$$

## 3.2 Backpropagation Algorithm (BP)

The back propagation (BP) algorithm is used to adjust all the weights. The inputs for BP are the output error and the established cell structure that contains all the setup parameters.

### 3.2.1 BP for Forecasting Model

In forecasting model, the output error at current time step *t* is the mean square error between current output and target value shown in (21)

$$E = \frac{1}{2}(y_t - Out_t)^2 \tag{21}$$

where $y_t$ and $Out_t$ are the target and output value respectively. The BP workflow diagram is proposed in Fig. 3.5, where three-time steps represented as green blocks are required to illustrate one BP procedure.

Starting from right to left, that is the reverse direction of time flow in DB-LSTM, the gradient of next cell state ($\frac{\partial E}{\partial C_{next}}$), current and next hidden state ($\frac{\partial E}{\partial h_t}$ *and* $\frac{\partial E}{\partial h_{next}}$) are inputted into the BP process at time step t. Since the output value $Out_t$ is set the same as hidden state $h_t$, the gradient of current hidden state can be expressed as

$$\frac{\partial E}{\partial h_t} = \frac{\partial E}{\partial Out_t} = y_t - Out_t \tag{22}$$

and the overall gradient to be propagated is

$$\frac{\partial E}{\partial h_t} = \frac{\partial E}{\partial h_t} + \frac{\partial E}{\partial h_{next}} \tag{23}$$

Next, (24) and (25) worked out the gradient with respect to output gate $o_t$ and output sum $o_s$ by partial differentiation.

$$\frac{\partial E}{\partial o_t} = \frac{\partial E}{\partial h_t} \cdot \frac{\partial h_t}{\partial o_t} = \frac{\partial E}{\partial h_t} \odot \tanh(c_t) \tag{24}$$

$$\frac{\partial E}{\partial o_s} = \frac{\partial E}{\partial o_t} \cdot \frac{\partial o_t}{\partial o_s} = \frac{\partial E}{\partial h_t} \odot \tanh(c_t) \odot \frac{d\sigma(o_s)}{do_s} \tag{25}$$

where $\frac{d\sigma(o_s)}{do_s}$ is the derivative of sigmoid function regarding to $o_s$, whose general equation is shown in (26)

$$\frac{d\sigma(x)}{dx} = x \odot (1 - x) \tag{26}$$

Referring to Fig. 3.5, the gradient of current cell state $\frac{\partial E}{\partial c_t}$ is contributed by three directions including the gradient of next cell state, current hidden state and current output gate, (27) – (31) explained the detailed computation.

$$\frac{\partial E}{\partial c_t} = \frac{\partial E}{\partial h_t} \cdot \frac{\partial h_t}{\partial c_t} \tag{27}$$

$$\frac{\partial h_t}{\partial c_t} = \frac{d\tanh(c_t)}{dc_t} \odot o_t + \tanh(c_t) \odot \frac{\partial o_t}{\partial c_t} \tag{28}$$

$$\frac{\partial o_t}{\partial c_t} = \frac{\partial o_t}{\partial o_s} \cdot \frac{\partial o_s}{\partial c_t} = c_o^T \times \frac{d\sigma(o_s)}{do_s} \tag{29}$$

where $\frac{d\tanh(c_t)}{dc_t}$ in (28) is the derivative of hyperbolic tangent function regarding to $c_t$, whose general expression is shown in (30)

$$\frac{d\tanh(x)}{dx} = 1 - (\tanh x \odot \tanh x) \tag{30}$$

and the overall gradient of $c_t$ is

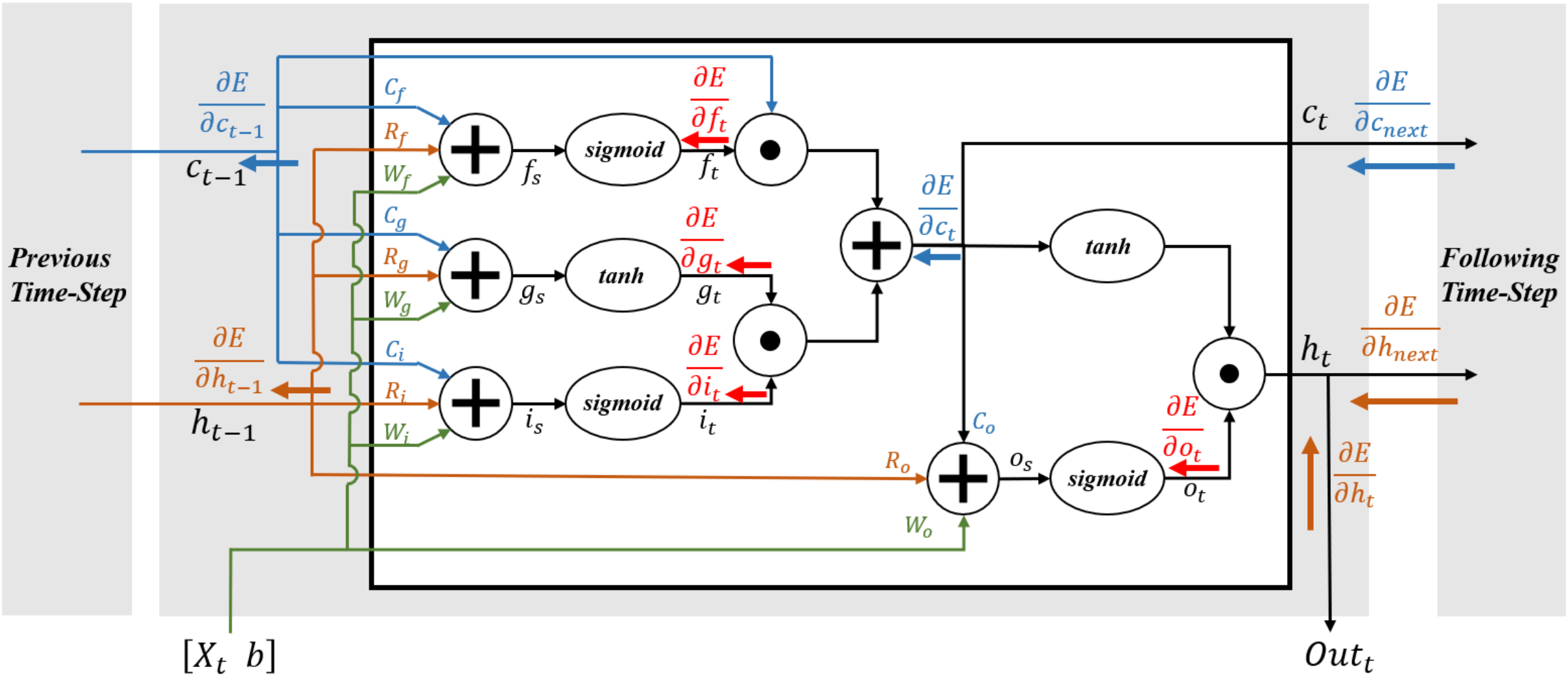

**Figure 3.5:** BP algorithm for forecasting model.

$$\frac{\partial E}{\partial c_t} = \frac{\partial E}{\partial c_t} + \frac{\partial E}{\partial c_{next}} \tag{31}$$

With the result of gradient of $c_t$, the gradient of forget, generate, and input gate can be derived by (32) – (34),

$$\frac{\partial E}{\partial f_t} = \frac{\partial E}{\partial c_t} \cdot \frac{\partial c_t}{\partial f_t} = \frac{\partial E}{\partial c_t} \odot c_{t-1} \tag{32}$$

$$\frac{\partial E}{\partial g_t} = \frac{\partial E}{\partial c_t} \cdot \frac{\partial c_t}{\partial g_t} = \frac{\partial E}{\partial c_t} \odot i_t \tag{33}$$

$$\frac{\partial E}{\partial i_t} = \frac{\partial E}{\partial c_t} \cdot \frac{\partial c_t}{\partial i_t} = \frac{\partial E}{\partial c_t} \odot g_t \tag{34}$$

and their gate sum can be further attained by (35) – (37).

$$\frac{\partial E}{\partial f_s} = \frac{\partial E}{\partial f_t} \cdot \frac{\partial f_t}{\partial f_s} = \frac{\partial E}{\partial c_t} \odot c_{t-1} \odot \frac{d\sigma(f_s)}{df_s} \tag{35}$$

$$\frac{\partial E}{\partial g_s} = \frac{\partial E}{\partial g_t} \cdot \frac{\partial g_t}{\partial g_s} = \frac{\partial E}{\partial c_t} \odot i_t \odot \frac{d\tanh(g_s)}{dg_s} \tag{36}$$

$$\frac{\partial E}{\partial i_s} = \frac{\partial E}{\partial i_t} \cdot \frac{\partial i_t}{\partial i_s} = \frac{\partial E}{\partial c_t} \odot g_t \odot \frac{d\sigma(i_s)}{di_s} \tag{37}$$

At this point, all gradients with respect to the DB-LSTM cell parameters at time step *t* have been yielded. The following tasks are propagating the gradient to previous time step as well as generating the weight gradient. Both previous hidden state and cell state gradients are connected with four data paths within the DB-LSTM cell, whose outcomes are shown in (38) and (39) respectively.

$$\begin{aligned}\frac{\partial E}{\partial h_{t-1}} &= \frac{\partial E}{\partial f_s} \cdot \frac{\partial f_s}{\partial h_{t-1}} + \frac{\partial E}{\partial g_s} \cdot \frac{\partial g_s}{\partial h_{t-1}} + \frac{\partial E}{\partial i_s} \cdot \frac{\partial i_s}{\partial h_{t-1}} + \frac{\partial E}{\partial o_s} \cdot \frac{\partial o_s}{\partial h_{t-1}} \\ &= R_f^T \times \frac{\partial E}{\partial f_s} + R_g^T \times \frac{\partial E}{\partial g_s} + R_i^T \times \frac{\partial E}{\partial i_s} + R_o^T \times \frac{\partial E}{\partial o_s}\end{aligned} \tag{38}$$

$$\frac{\partial E}{\partial c_{t-1}} = \frac{\partial E}{\partial c_t} \cdot \frac{\partial c_t}{\partial c_{t-1}} \tag{39}$$

where

$$\frac{\partial c_t}{\partial c_{t-1}} = f_t + c_{t-1} \odot \frac{\partial f_t}{\partial c_{t-1}} + i_t \odot \frac{\partial g_t}{\partial c_{t-1}} + g_t \odot \frac{\partial i_t}{\partial c_{t-1}}$$

$$\frac{\partial f_t}{\partial c_{t-1}} = \frac{\partial f_t}{\partial f_s} \cdot \frac{\partial f_s}{\partial c_{t-1}} = C_f^T \times \frac{d\sigma(f_s)}{df_s}$$

$$\frac{\partial g_t}{\partial c_{t-1}} = \frac{\partial g_t}{\partial g_s} \cdot \frac{\partial g_s}{\partial c_{t-1}} = C_g^T \times \frac{d\tanh(g_s)}{dg_s}$$

$$\frac{\partial i_t}{\partial c_{t-1}} = \frac{\partial i_t}{\partial i_s} \cdot \frac{\partial i_s}{\partial c_{t-1}} = C_i^T \times \frac{d\sigma(i_s)}{di_s}$$

On the other hand, three groups of weight gradient at time step t can be derived involving (25) and (35) – (37), shown in (40) – (43).

$$\frac{\partial E}{\partial W_{f,g,i,o}} = \frac{\partial E}{\partial f,g,i,o_s} \times [X_t \;\; b]^T \tag{40}$$

$$\frac{\partial E}{\partial R_{f,g,i,o}} = \frac{\partial E}{\partial f,g,i,o_s} \times h_{t-1}^T \tag{41}$$

$$\frac{\partial E}{\partial C_{f,g,i}} = \frac{\partial E}{\partial f,g,i_s} \times c_{t-1}^T \tag{42}$$

$$\frac{\partial E}{\partial C_o} = \frac{\partial E}{\partial o_s} \times c_t^T \tag{43}$$

The above equations for BP in forecasting will be calculated for a total time steps of k in each training iteration. At the end of each training iteration, the weights will be updated using the accumulated cost gradient with respect to each weight for all time steps. With a learning rate of $\eta$, the updating for all three groups of weight can be expressed by (44) for the program to run to the next epoch

$$\delta W,R,C = \sum_k \frac{\partial E}{\partial W,R,C}$$
$$W,R,C = W,R,C - \eta \cdot \delta W,R,C \tag{44}$$

### 3.2.2 BP for Classification Model

As for classification model, the output error at time step *t* is expressed as the cross-entropy loss function shown in (45)

$$E = -\sum_{i=1}^{n} y_i \log(p_i) \tag{45}$$

where n stands for the number of classes, $y_i$ is the truth label and $p_i$ is the SoftMax probability for the i$^{th}$ class. Since Chapter 3.1 has introduced the difference between forecasting and classification models, the BP process of classification model is proposed in Fig. 3.6, whose green block stands for BP computation of DB-LSTM layer for all k time steps. The cross-entropy loss will be propagated to FC layer to obtain the gradient with respect to the last hidden state of DB-LSTM layer shown in (45).

$$\begin{aligned}\frac{\partial E}{\partial Out_j} &= -\sum_{i=1}^{n} y_i \frac{\partial \log(p_i)}{\partial Out_j} \\ &= -\sum_{i=1}^{n} y_i \frac{\partial \log(p_i)}{\partial p_i} \times \frac{\partial p_i}{\partial Out_j} \\ &= -\sum_{i=1}^{n} y_i \frac{1}{p_i} \times \frac{\partial p_i}{\partial Out_j}\end{aligned} \tag{45}$$

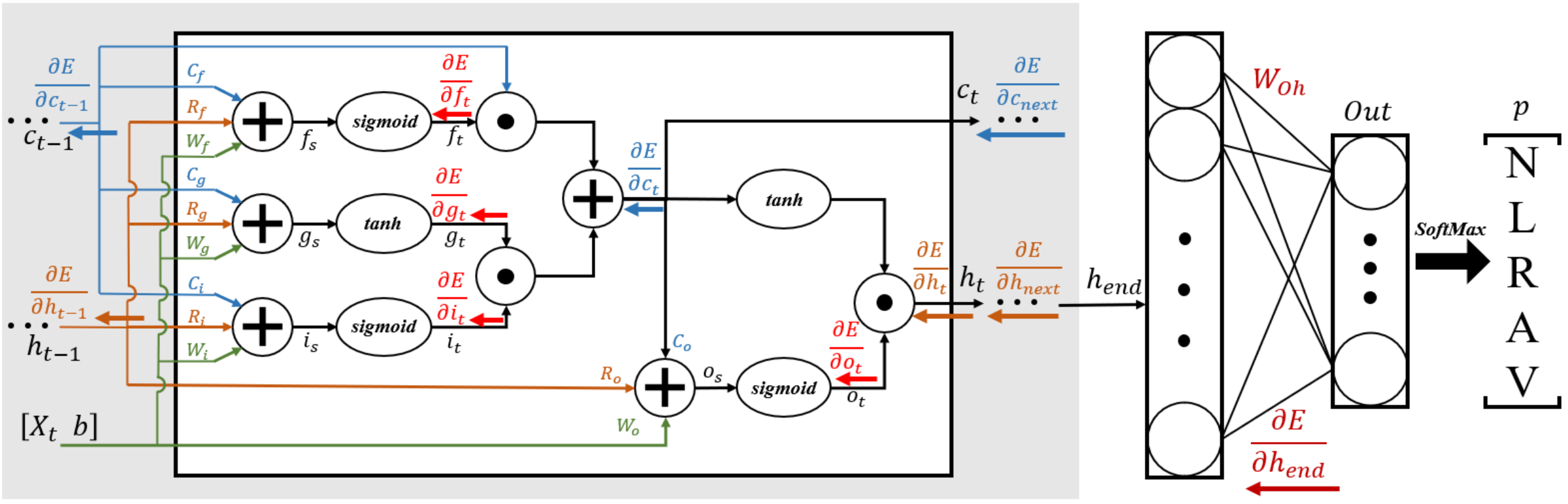

**Figure 3.6:** BP algorithm for classification model.

where $\frac{\partial p_i}{\partial Out_j}$ is the derivative of SoftMax function obtained in (46)

$$\frac{\partial p_i}{\partial Out_j} = \begin{cases} p_i(1-p_j), & i = j \\ -p_i \cdot p_j, & i \neq j \end{cases} \tag{46}$$

As a result, the gradient with respect to output class is

$$\frac{\partial E}{\partial Out_j} = p_j - y_j$$

and the gradient regarding to final hidden state of DB-LSTM layer and the weight gradient of FC layer are expressed in (47) and (48) respectively.

$$\frac{\partial E}{\partial h_{end}} = \frac{\partial E}{\partial Out} \cdot \frac{\partial Out}{\partial h_{end}} = W_{Oh}^T \times \frac{\partial E}{\partial Out} \tag{47}$$

$$\frac{\partial E}{\partial W_{Oh}} = \frac{\partial E}{\partial Out} \cdot \frac{\partial Out}{\partial W_{Oh}} = \frac{\partial E}{\partial Out} \times Out^T \tag{48}$$

Comparing to the BP process in Fig. 3.5, only the last time step of DB-LSTM receives the output error gradient from the next fully connected layer, while all other hidden states only do recursing. Thus, the gradient of hidden state for all k time steps can be modified to (49) based on (37). By obtaining the gradient of hidden state in (49), the following BP algorithm is same as (33) – (43) to attain the update of three groups of weight. Similarly, the update of $W_{Oh}$ follows the same rule as (43) with learning rate of $\eta$.

$$\frac{\partial E}{\partial h_t} = \begin{cases} \frac{\partial E}{\partial h_{end}}, & t = k \\ \frac{\partial E}{\partial h_{next}}, & t \neq k \end{cases} \tag{49}$$

### 3.3 Fix-point Weight Training

The weights in the AI algorithms are normally represented in FP32. However, as the goal of this work is to implement the algorithm in portable device, it is important to quantize the weights. So that they can fit into the limit hardware resources of edge device with acceptable accuracy degradations. The weights are mapped to fixed-point numbers, which are quantized into signed $2 \times 2^n$ states including negative values, where n is the bit length. The quantization procedure is to firstly find the maximum and minimum numbers in that weight matrix in absolute value, denoted as $w_{max}$ and $w_{min}$, respectively. They create the boundary of states, and $w_{min}$ is set as the first state $s_1$. The interval between these states is given by

$$\Delta w = \frac{w_{max} - w_{min}}{2^n - 1} \tag{50}$$

where the classify interval is half of $\Delta w$. For each weight entry $w_i$ within that weight matrix, the mapping method obeys

$$w_i = s_{j+1}, \qquad if\ |w_i - (s_1 + j \times \Delta w)| \leq \frac{\Delta w}{2} \tag{51}$$

where $j = 0, 1, 2, \dots 2^n - 1$. Fig. 3.7 compares the training accuracy with different weight precisions of FP32 and fixed bit length quantization in forecasting and classification training, respectively. According to Fig. 3.7(a), it can be observed that floating weights achieved the fastest convergence and has high accuracy. On the other hand, with shorter fixed weight precision, the convergence speed becomes slower, though there is no degradation of accuracy. This is because the result is based on normal ECG dataset whose QRS sequence varies periodically. If an abnormal dataset is tested, the accuracy will decline as indicated in Table III, whose details will be discussed in Chapter 4. Fixed-point weight precision with the best accuracy is also selected for comparison with the floating case. Since the weights in fixed-point are quantized into finite numbers, the initial random weight can significantly affect the converge direction in training the model, which increased the failure rate to around 5% for INT1 precision. While in classification task in Fig. 3.7(b), INT4 weight precision obtained similar accuracy and loss as floating. On the other hand, with shorter fixed-point weight precision, INT3 for instance, the convergence speed became slower, and the degradation of accuracy is around 6%. That is because finite choice of weights will limit the computing sparsity and result indistinguishable outputs.

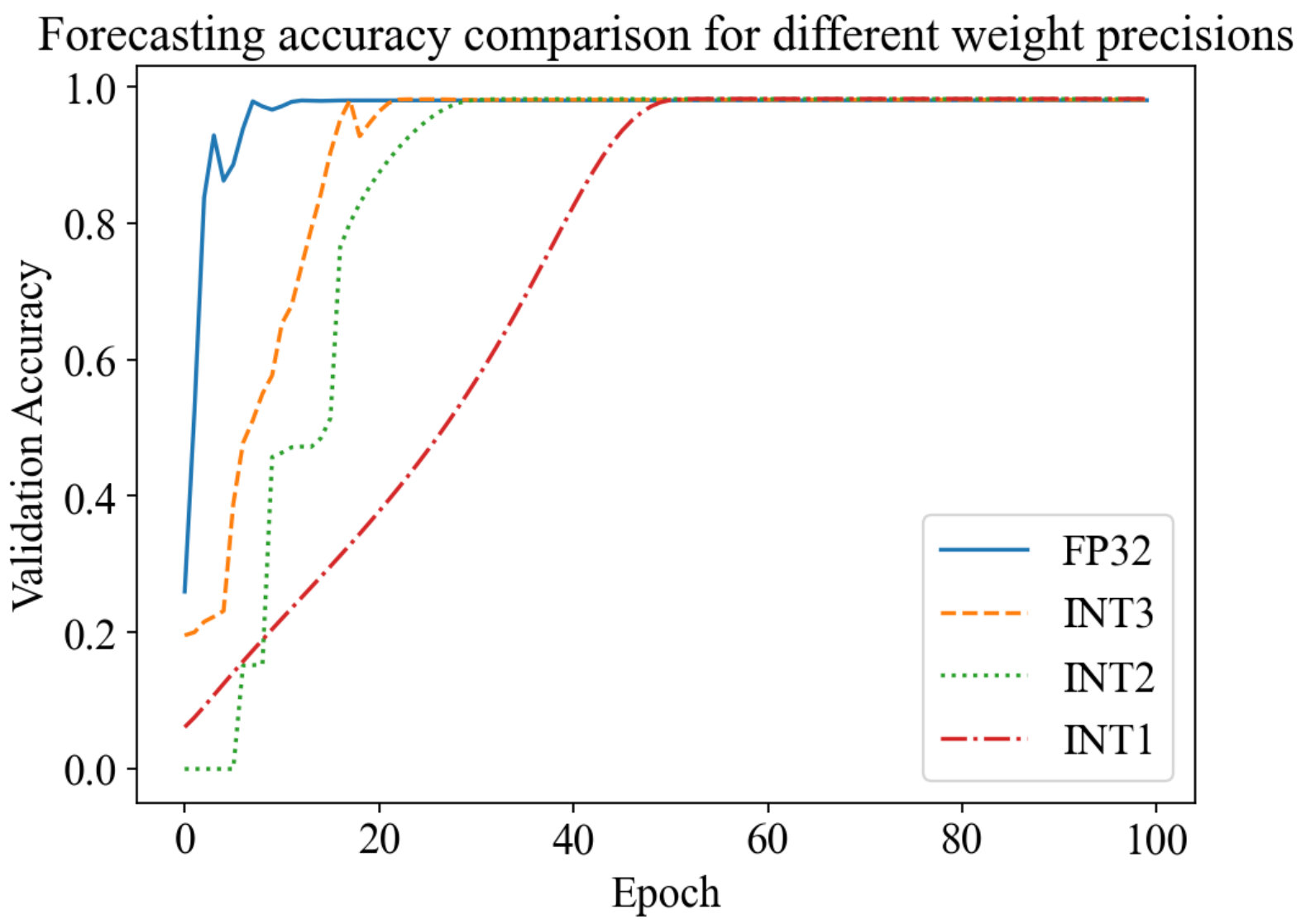

(a)

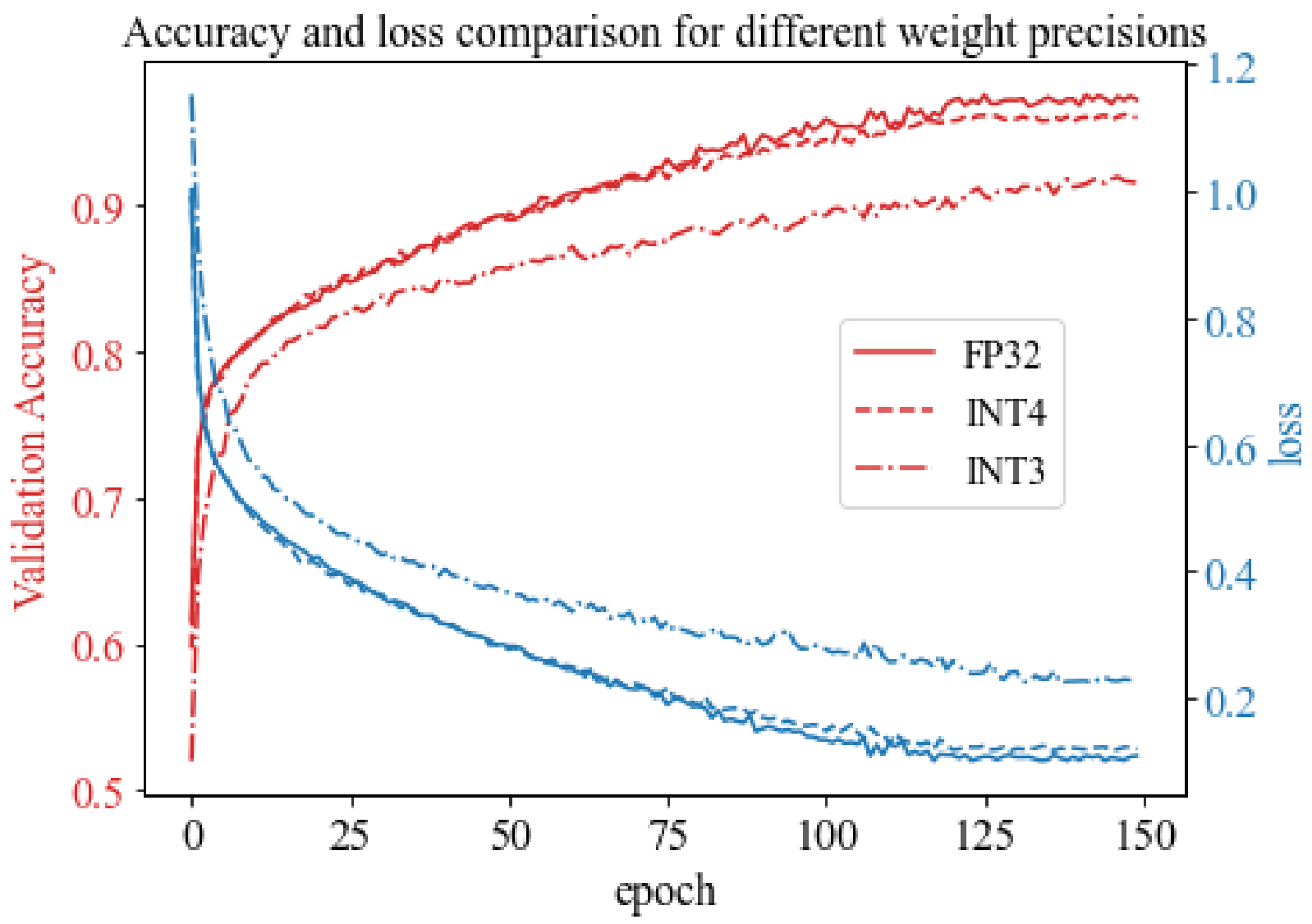

(b)

**Figure 3.7:** Comparison of training process with different weight resolution in: (a) forecasting model and (b) classification model.

TABLE III MULTI DATASET PERFORMANCE COMPARSION

| Database | Category | Total Length | Prediction Ability | Average Accuracy | Average NMSE |
|---|---|---|---|---|---|
| MIT-BIH | N | 420 min | 1 HB | 98.86% | 0.00026 |
| MIT-BIH | V | 42 min | 2 HBs | 95.31% | 0.00049 |
| MIT-BIH | R | 41 min | 1 HB | 94.96% | 0.00058 |
| MIT-BIH | L | 45 min | 1 HB | 96.79% | 0.00061 |
| MIT-BIH | P | 39 min | 1 HB | 97.56% | 0.00025 |
| INCART | N | 30 min | 1 HB | 94.53% | 0.00083 |
| INCART | V | 30 min | 2 HBs | 93.53% | 0.00086 |

# Chapter 4

# Results

## 4.1 Introduction

Both ECG forecasting and classification tasks are applied to verify the robustness of DB-LSTM network. To classify ECG singles efficiently, both patients and medical doctors can notice the type of cardiac diseases fast and apply necessary treatments in time. When the patient is alone and outbreak illness accidently, even if the patient is unconscious that cannot talk about the symptom, this portable classification system can guide the physician to implement medication punctually. Besides just classifying signals, this system can further forecast the waveform of following heartbeats since every patient's cardiac disease may differ although belonging to the same type of illness. Thus, it is also valuable to show the following heartbeat waveform to doctors to prescribe suitable treatment.

## 4.2 DB-LSTM Cell Performance

### 4.2.1 Theoretical Model Performance

The theoretical performance shown in Fig. 4.1 is a comparison of loss between traditional LSTM [26] and the proposed DB-LSTM model with 32-bit precision weight. The training data is a multi-sequence to sequence prediction. Here, the input is a 200×15 matrix and the output is a 200×1 vector, where 200 is the sequence length. It can be observed that the proposed method achieved better performance than the original LSTM in both final loss and training stability. Also, this model attained lower initial loss but a rapid loss reduction, which is due to the larger memory size of DB-LSTM cell structure. Moreover, the dynamic bias of LSTM can make the training procedure be stable. Thus, the result will not be reflected by the oscillation of descending gradient.

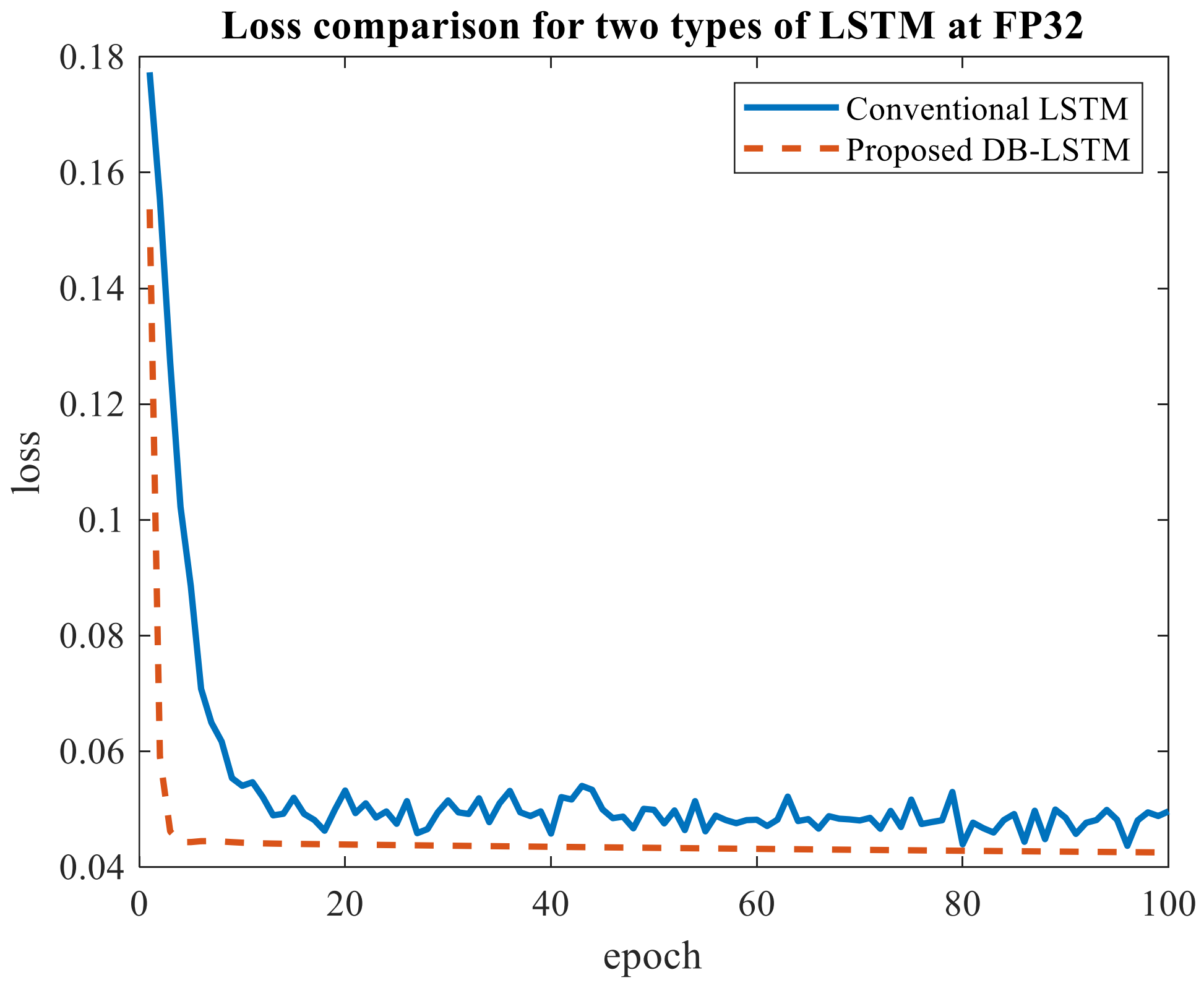

**Figure 4.1:** Comparison of training loss at default 32-bit weight.

### 4.2.2 Application on Real Life Problem

To verify the performance of this DB-LSTM model solving practical problem, a time series forecasting for air passengers [28] is evaluated.

In this problem, a time series of monthly traveller numbers is fed into the model, while the label output is the same sequence but delay by one month. The aim is to have the model forecasts the next month's passenger volume by providing the current data, and then repeat this procedure to predict for as many months as it can. Fig. 4.2(a) shows the accuracy and loss diagram in three precision weights to train the model with prior 109 months' data, whereby the inputs are from 1 to 108 and the label output is from 2 to 109. The accuracy of this model reached 95.49%, 93.97%, 94.96% for 32bit, fix 1-, and 2-point weights, correspondingly. Fig. 4.2(b) shows the difference between real and predict passengers in the next 34 months with three types of weight precision, from which the forecasting results for the next two years indicated a small diversity. The mean square error for 32-bit, 2-bit and 1-bit weight validation are 156.34, 168.77 and 271.26, respectively using the same training parameters.

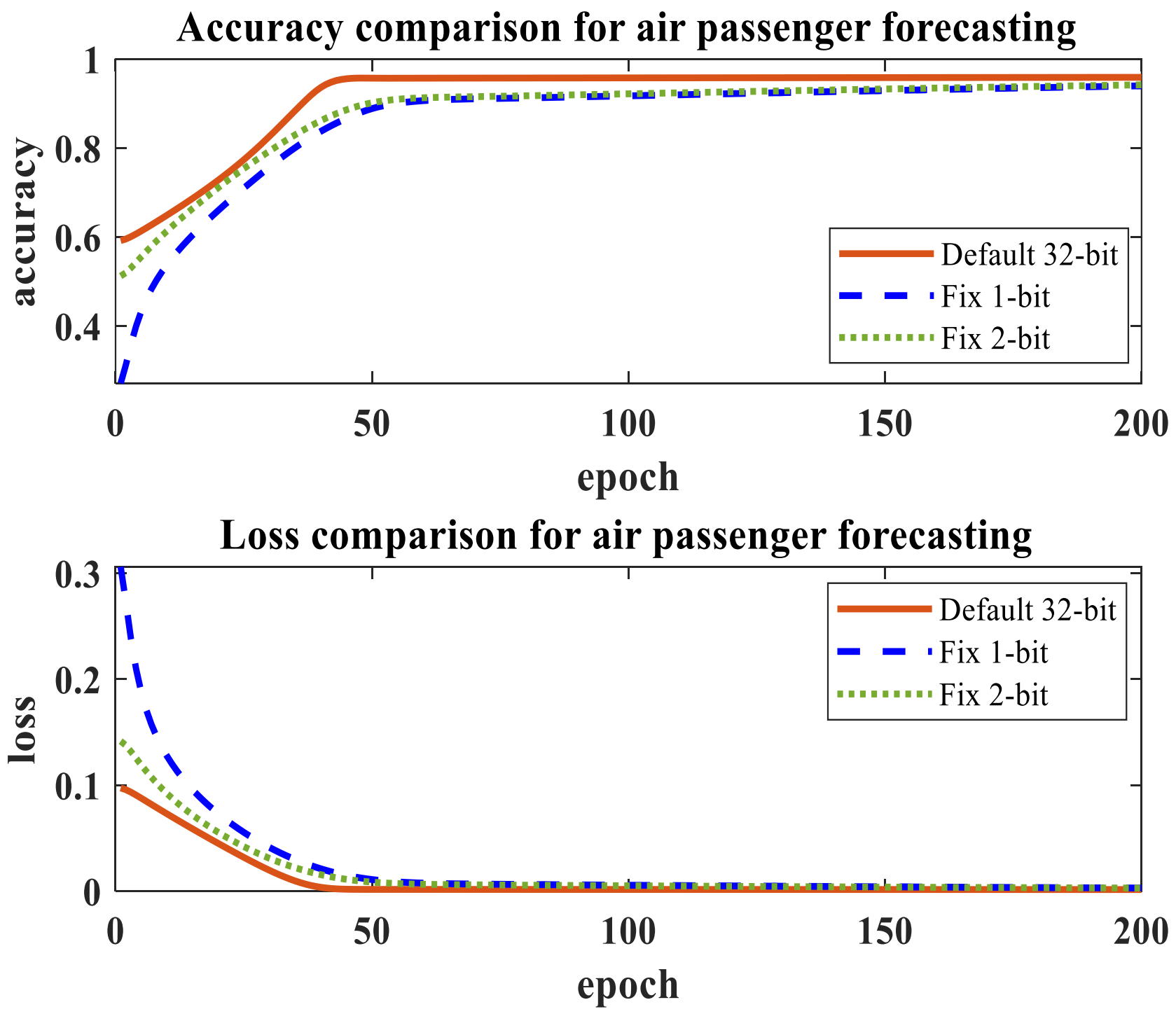

(a)

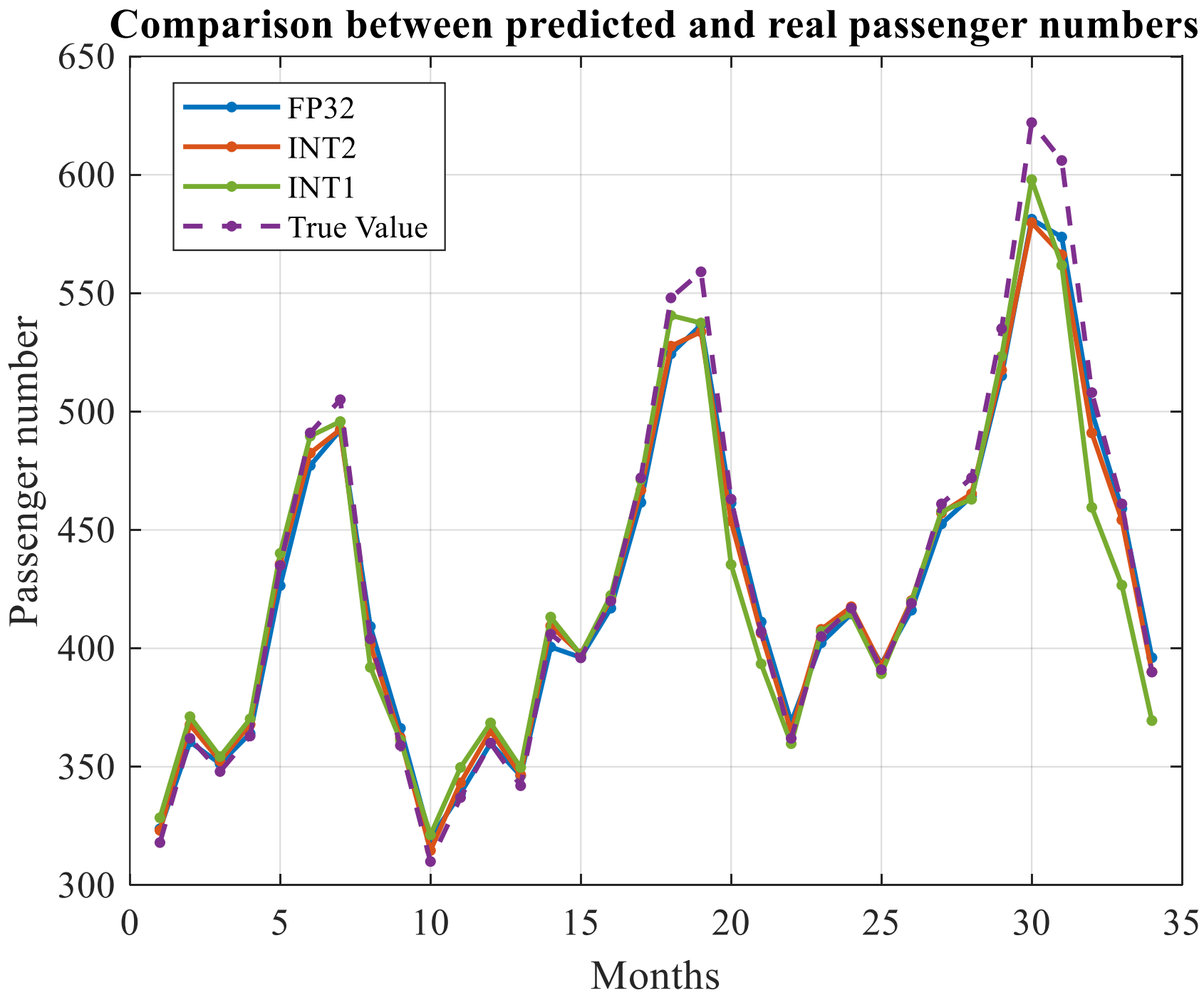

(b)

**Figure 4.2:** Comparison (a) of training accuracies and losses for air passenger forecasting (b) between air passenger number true value and forecast results, with different weight quantization accuracies.

Additionally, this model can yield satisfactory results in a shorter training time for one-dimension time series forecasting. This implies a higher probability for online training on edge devices. In the 2-bit precision experiment, it takes 200 epochs to reach a 94.96% accuracy and around 96% accuracy for 1000 epochs. This is of a better result than [13] which required 800 epochs training using the same dataset. A similar ELSTM [29] was presented with less gates for time series forecasting and attained 90.89% accuracy after training for 20,000 epochs. It saves on hardware cost but is only suitable for pre-training due to its long training period. Compared with the 32-bit full-precision baseline, the 1-bit and 2-bit quantization achieved approximately 16.7% average memory saving and 9.8% inference acceleration on CPUs, with only a reasonable loss in the accuracy.

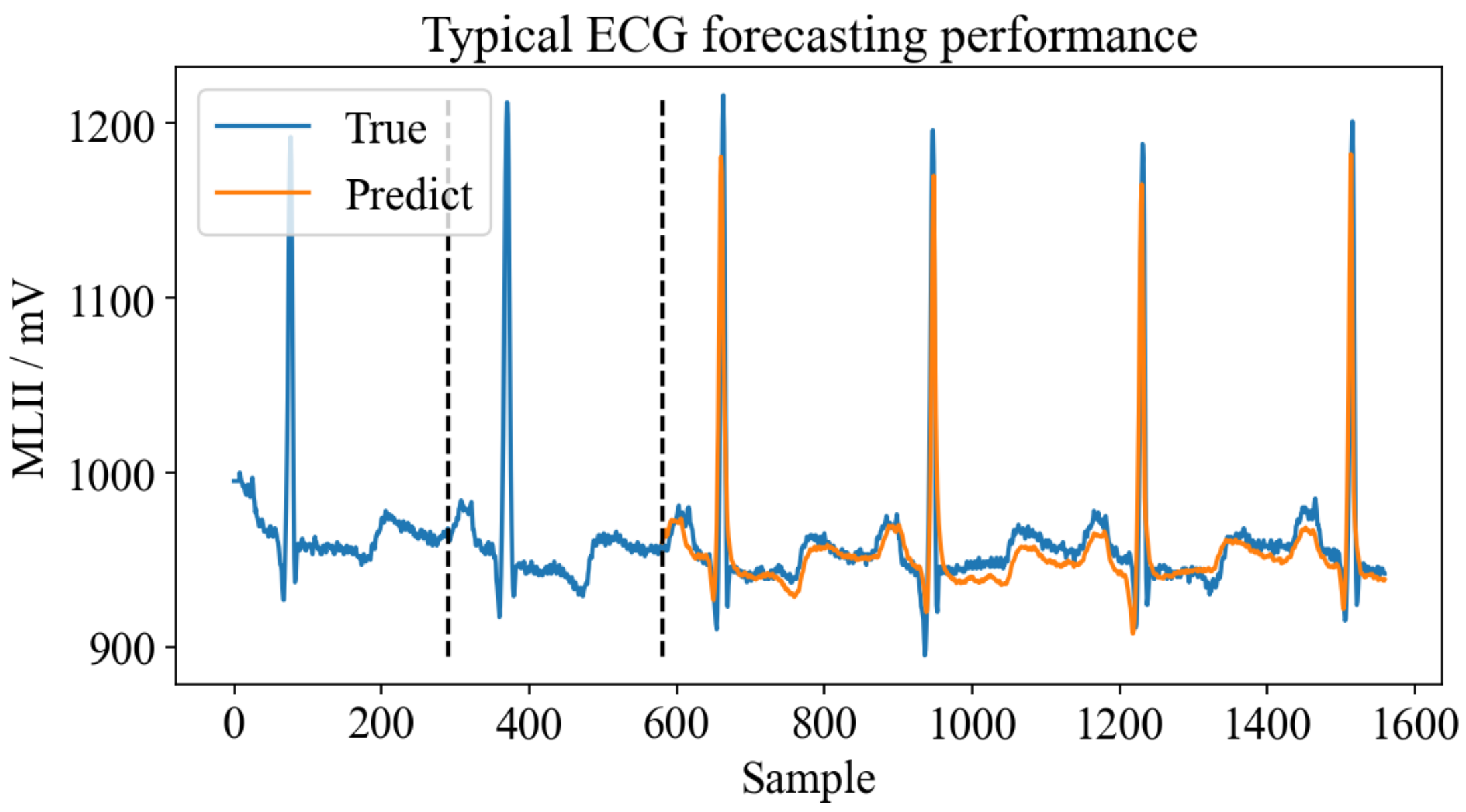

(a)

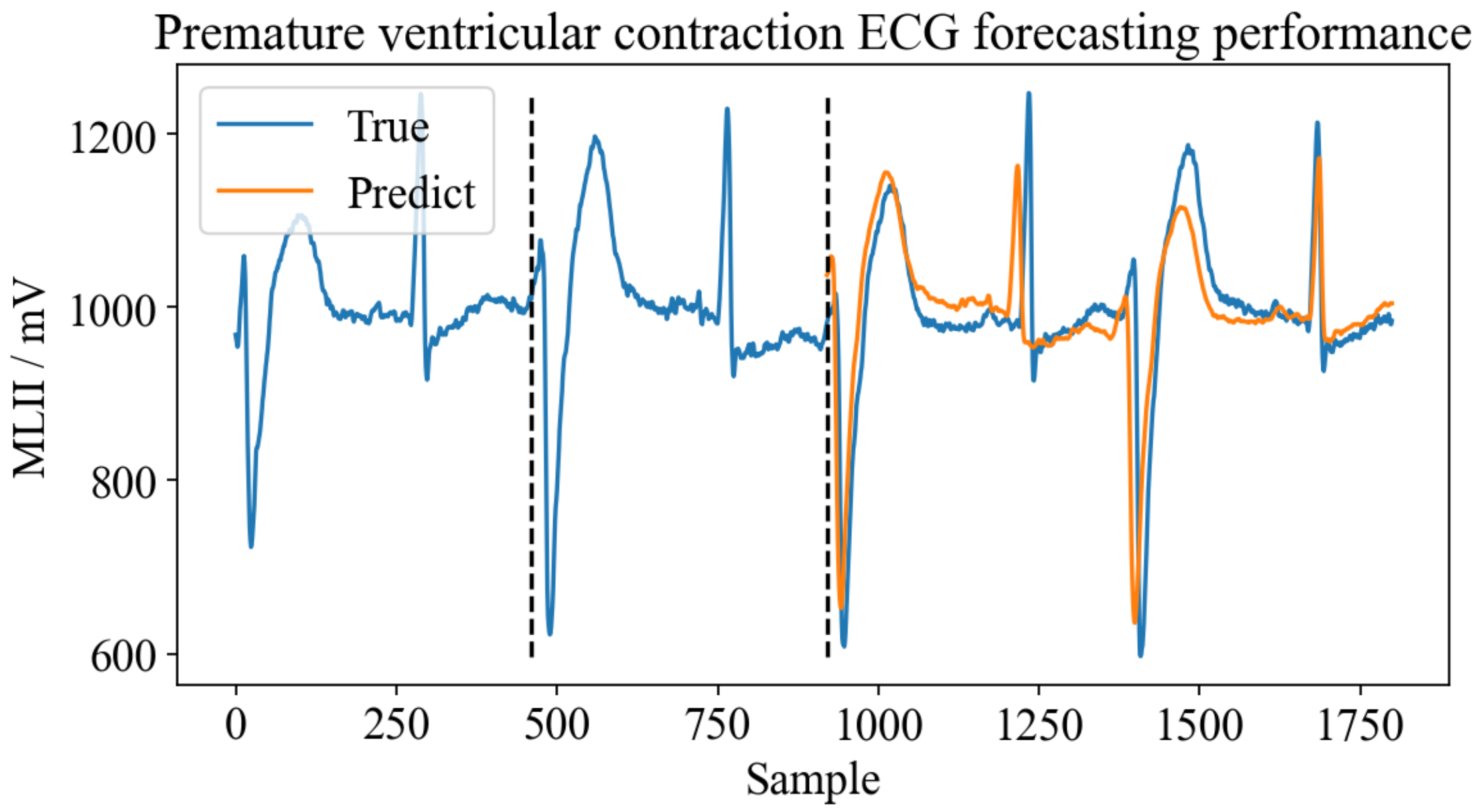

(b)

**Figure 4.3:** Forecasting results of: (a) normal and (b) PVC ECG waveforms.

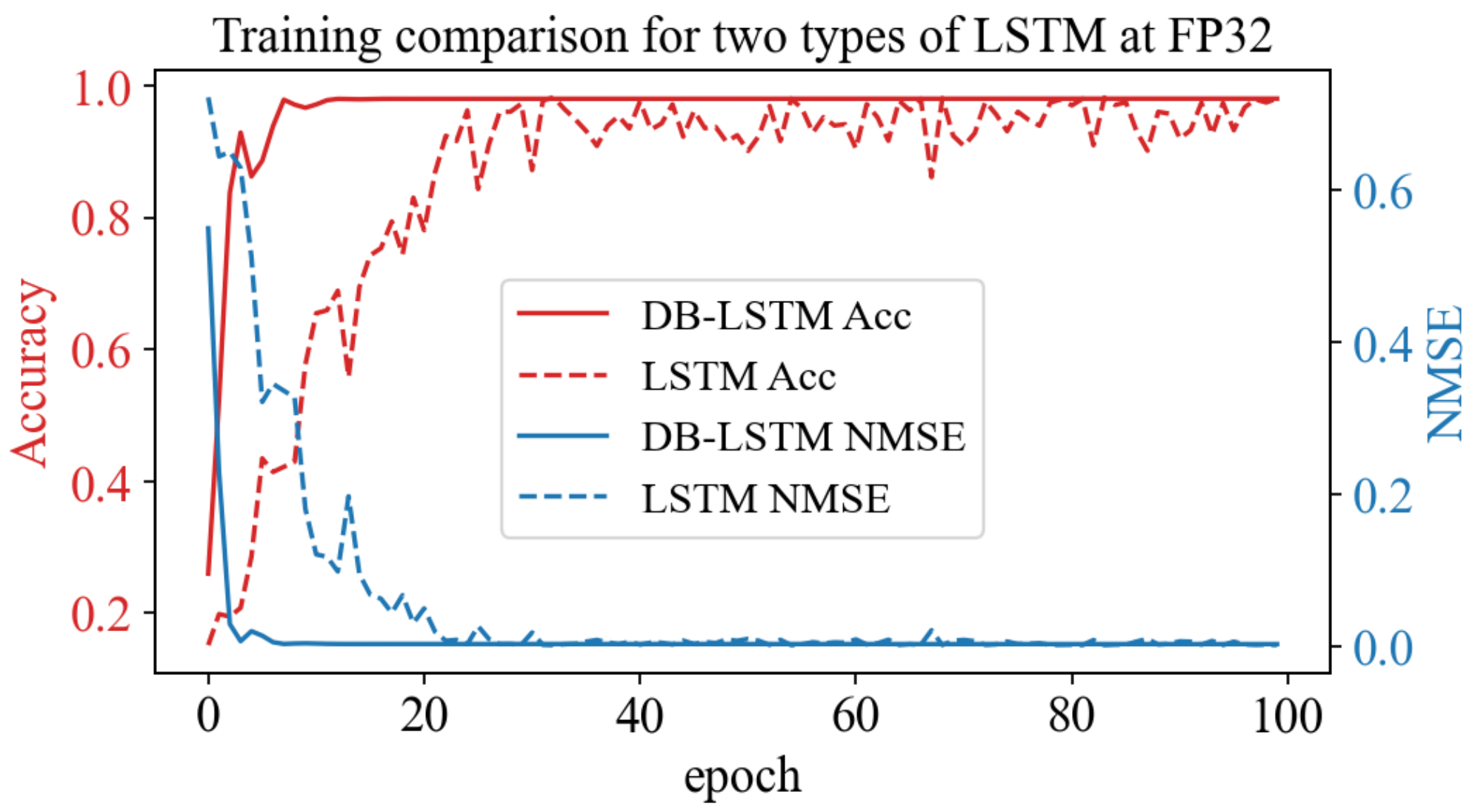

**Figure 4.4:** Training comparison between DB-LSTM and conventional LSTM.

### 4.3 Results for ECG Forecasting

The following study focused on the forecasting for the normal ECG and premature ventricular contraction (PVC) ECG. For normal ECG signal, each heartbeat is similar and regular. Hence the PQRST peaks are easy to differentiate. By selecting the delay period of 280 samples which are the heartbeat rates of the ECG signal, DB-LSTM can be trained within 560 samples for forecasting of consecutive heartbeats. Fig. 4.3(a) illustrates the time series prediction result for dataset 101 in MIT-BIH database, which is a typical Normal ECG signal. With only 100 epochs training, DB-LSTM can realize trained and validation accuracy of 99.23% and 98.86%, respectively; the RMSE of train and validation set are 0.00021 and 0.00026, respectively.

As for ECG associated with disease information, each type of ECG wave provides unique characteristics, which can be memorized by DB-LSTM. Thus, an abnormal ECG signal such as the Premature Ventricular Contraction (PVC) is chosen to test the creditability of DB-LSTM. PVC [30] had an advanced occurrence of QRS complexes than the normally conducted heartbeat. Also, the duration of PVC is longer than the typical one. The direction of ST segment and T wave is opposite to the main wave direction of QRS complexes. The full cycle of PVC is two heartbeats including one PVC and one normal contraction. Thus, the DB-LSTM is able to forecast the following two heartbeats ECG wave. With same model parameters of DB-LSTM, that is, 0.1 learning rate, 0.01 weight penalty, and 100 epochs, Fig. 4.3(b) states the forecasting result of dataset 200 in MIT-BIH database. With a delay duration of 460 samples, it reached a train and validation accuracy of 95.97% and 95.31%, respectively, where their RMSE being 0.00051 and 0.00049 respectively.

To verify the robustness of DB-LSTM model, different databases are tested to validate the performance of this model, whose results are tabulated in Table III. Two databases, MIT-BIH and INCART [31] with five categories of ECG signals are realized to conduct both the accuracy and Normalized Mean Square Error (NMSE) evaluation. All datasets are over 30 minutes so that the prediction results are reliable. For premature ventricular contraction, DB-LSTM needs to be trained for two heartbeats. With following two heartbeats forecast, whereas the other four types of ECG can be adequately trained with single heartbeat. Consequently, this DB-LSTM had verified its ability to deal with multi-variate ECG signals.

The training process of DB-LSTM is much faster and more stable than the conventional LSTM [26], whose comparison is shown in Fig. 4.4. The comparison dataset is chosen to be the 101 normal heartbeat ECG signal. With same learning rate, weight penalty and epochs, the DB-LSTM is associated with a lower initial loss and higher training accuracy than the conventional LSTM. Moreover, a faster and stabler training period can render the real AIoT devices more user efficient and accurate.

TABLE IV MULTI MODEL PERFORMANCE COMPARSION

| Topology | Number of NN Layer | Window Size | Number of Weight Parameters | Validation Accuracy | Weight Quantization |
|---|---|---|---|---|---|
| Deep CNN | 9 | 360 | 22,181 | 98.62% | FP32 |
| Deep CNN | 9 | 180 | 14,981 | 96.66% | FP32 |
| LSTM | 4 | 360 | 4,517 | 87.26% | FP32 |
| LSTM | 4 | 180 | 4,517 | 84.90% | FP32 |
| BiLSTM | 4 | 180 | 9,029 | 95.92% | FP32 |
| DB-LSTM | 4 | 180 | 8,608 | 91.8%, 96.2% & 97.5% | INT3, INT4 & FP32 |

### 4.4 Results for ECG Classification

The performance of the proposed algorithm is evaluated with MIT-BIH ECG database which includes 48 sets of 30-minute-long ECG data. Each dataset contains both normal and abnormal ECG features (N, L, R, A, V) shown in Fig. 1.1 recorded with a fixed sampling frequency of 360 Hz and 11-bit data resolution.

After pre-processing, the training dataset contains 20,000 samples with 4000 samples in each category and the validation dataset has 5000 samples with 1000 samples in each category. The output size of DB-LSTM cell is set as 32, thus the total number of training parameters is 8608. With 0.01 learning rate, 0.05 weight gradient, 0.01 weight penalty and cross-entropy loss function, the validation accuracy for five-category classification with FP32 weight precision is 97.5%, the corresponding confusion matrix is shown in Fig. 4.5(a). Furthermore, after weight being quantized into INT4 and INT3, the validation accuracy dropped to 96.2% and 91.8%, as shown in Fig. 4.5(b) and (c), respectively. For each confusion matrix, the horizontal axis stands for the predicted category and the vertical axis gives the actual label. For each entry in the matrix, the upper data is the number of predicted samples, while the lower one is the percentage out of the total validation dataset. The summation of diagonal entries is the total validation accuracy.

Various Deep CNN, conventional LSTM and bidirectional LSTM models are built to compare the performance, the results are tabulated in Table IV. Different window sizes are used to characterize the influence on accuracy and training parameters. With the same deep CNN structure, double window size will result in almost double in total training parameters, but only 2% improvement of accuracy. On the other hand, the training parameters of LSTM will not be affected by the variance of window size, which only depends on the input and output size. Comparing to conventional LSTM, bidirectional LSTM doubles the number of parameters and enhance the accuracy by 8.5%. The drawback of it is that bidirectional cannot perform continuous time monitoring, which yields a large latency in real-time classification. DB-LSTM yields short window size, continuous time processing, less training parameters and higher classification accuracy than other models. It reached similar classification result as Deep CNN with same window size, but a smaller number of neural network layers and 43.6% less of training parameters.

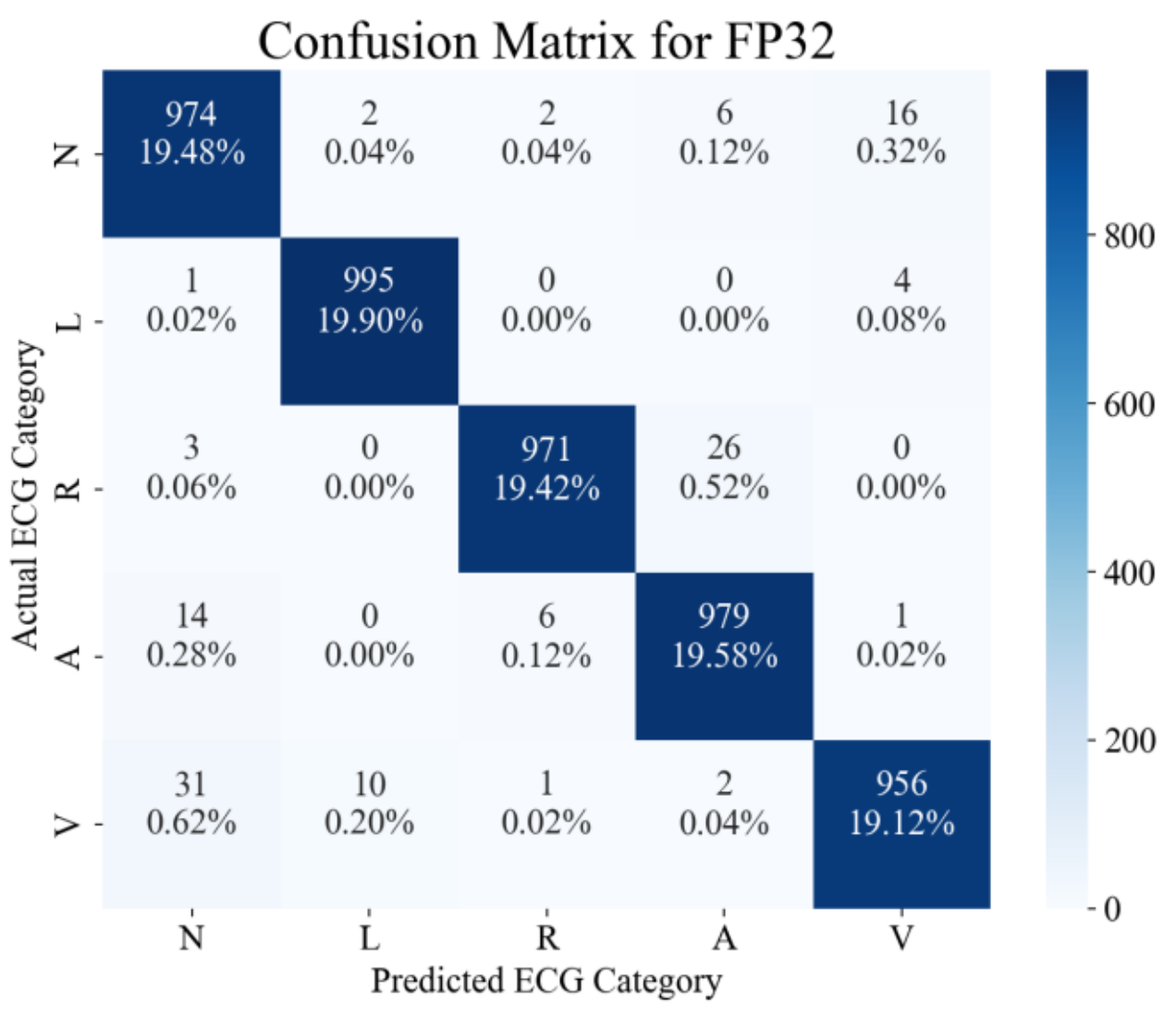

(a)

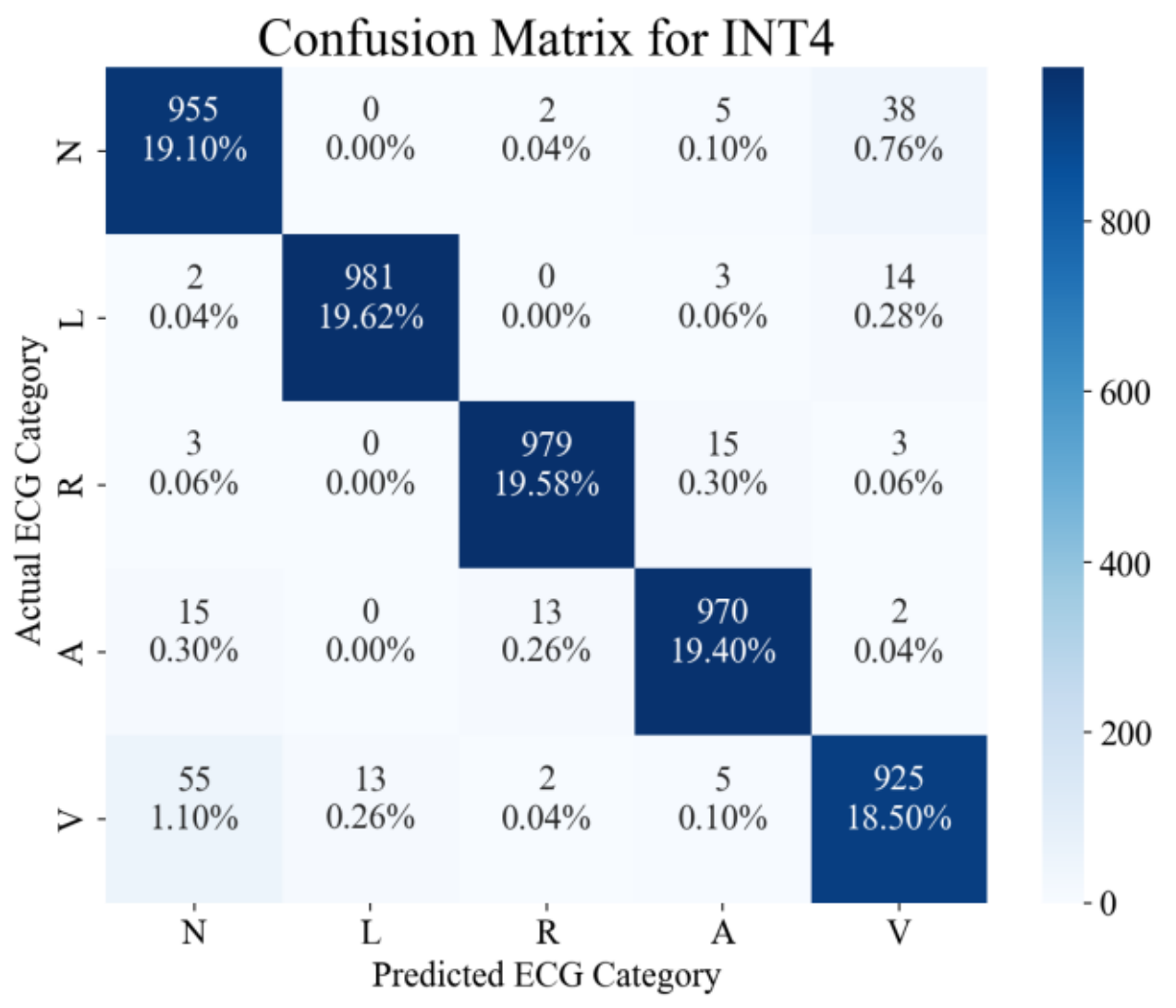

(b)

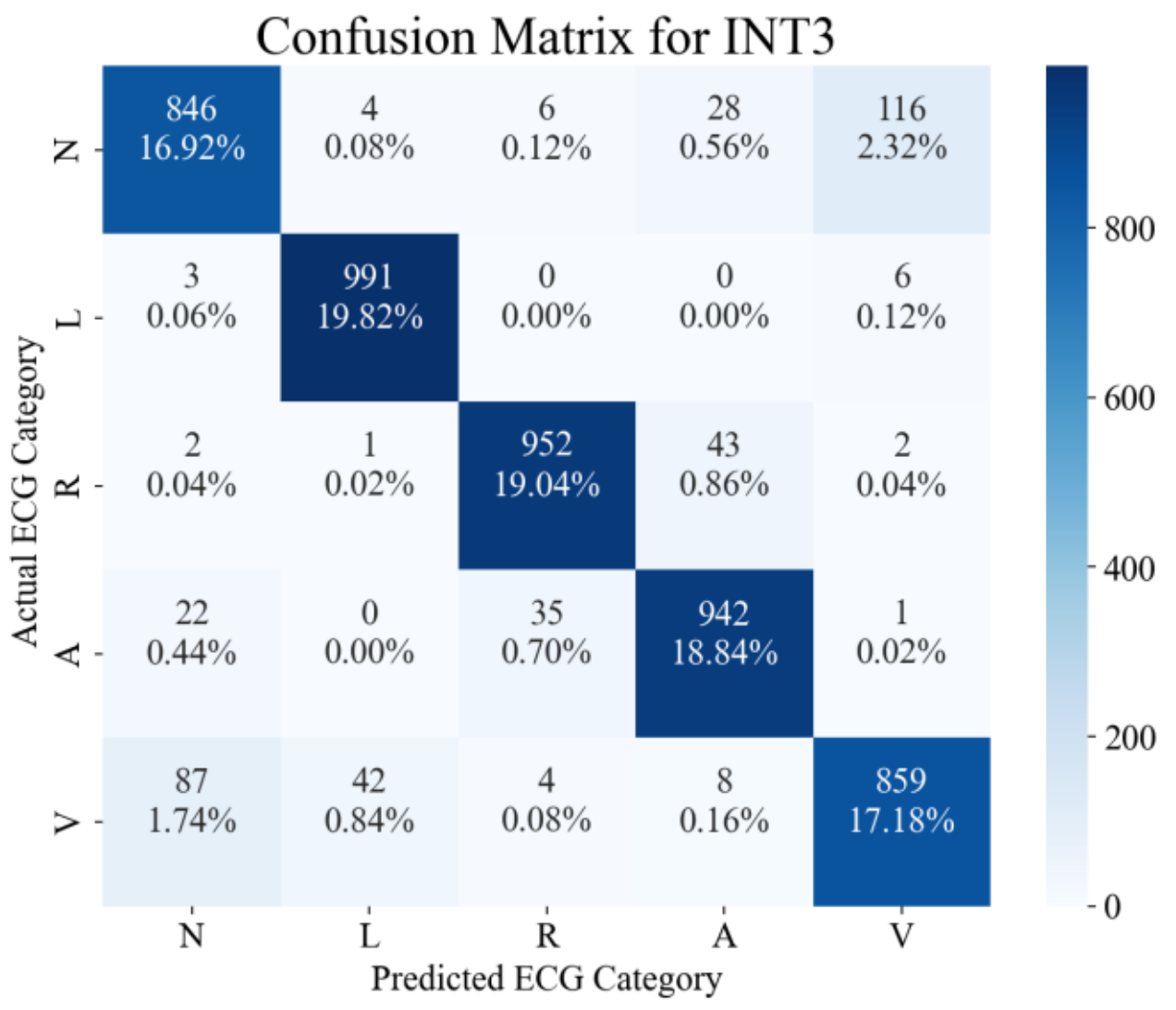

(c)

**Figure 4.5:** Classification confusion matrices for different weight precisions, (a) FP32, (b) INT4 and (c) INT3.

# Chapter 5

# Discussion

This section focuses on the comparison between this DB-LSTM system and other top-north systems to validate its strong effects on both forecasting and classification applications.

## 5.1 Discussion on ECG Forecasting

This DB-LSTM model is compared with other 1-dimentional time-series forecasting on bio signals, such as the ECG, Electroencephalogram (EEG) and surface electromyography (sEMG), as benchmarked in Table V.

A 500-sample forecasted system using Deep LSTM for EEG prediction had been reported in [32]. The model is structured by 5-layer NN, consisting of two LSTM layers and one fully connected layer. With 18 LSTM neurons for each layer, the total number of weights is 450. After 1280 samples, it achieved an accuracy of 88.9%. An online training LSTM [33] achieved low forecasting error but required long training samples, that could result in high latency when training on edge devices. A state-of-art application on sEMG real time forecasting stated in [34] demonstrated ability to forecast 25 seconds ahead with CNN. A stack of six dilated convolutional layers with two filters in size of length 4×64 is embedded in this model. Accompanied by a fully connected layer, the total learnable parameters are $6.4\times10^{6}$, which consumed large memory and high power in hardware implementation. This model is well trained, based on a total number of 108342 samples. It had conducted online monitoring of low back pain with 82.8% accuracy.

TABLE V BENCHMARK TABLE FOR FORECASTING

| | [16] | [17] | [18] | [32] | [33] | [34] | This Work |
|---|---|---|---|---|---|---|---|
| Topology | VMD + NN | PSR + NN | TS Fuzzy | Deep LSTM | ef-WMF-LSTM | CNN | DB-LSTM |
| Application | ECG | ECG | ECG | EEG | Alcoa | sEMG | ECG |
| Forecast Steps (Samples) | 1 | 1 | 1 | 500 | 1 | 5875 | 280 |
| Number of Layer | 3 | 3 | NA | 5 | 4 | 9 | 3 |
| No. of Weight Parameters | 180 | 300 | 7201* | 450* | 220 | $6.4\times10^{6}$* | 16 |
| Training Samples | 9×2160 | 9×2880 | 2500 | 1280 | 1000 | 108342 | 280 |
| Training Method | Pre-Training | Pre-Training | Pre-Training | Pre-Training | Online Training | Pre-Training | Online Training |
| RMSE | 0.0233 | 0.0423 | 0.0146 | NA | 0.0042 | NA | 0.0352 ± 0.0021 |
| Validation Accuracy | 98.8% | 97.3% | NA | 88.9% | NA | 82.8% | 98.1 ± 0.15% |
| Weight Quantization | FP32 | FP32 | FP32 | FP32 | FP32 | FP32 | INT3, INT2 and INT1 |

* Calculated based on information provided in paper.

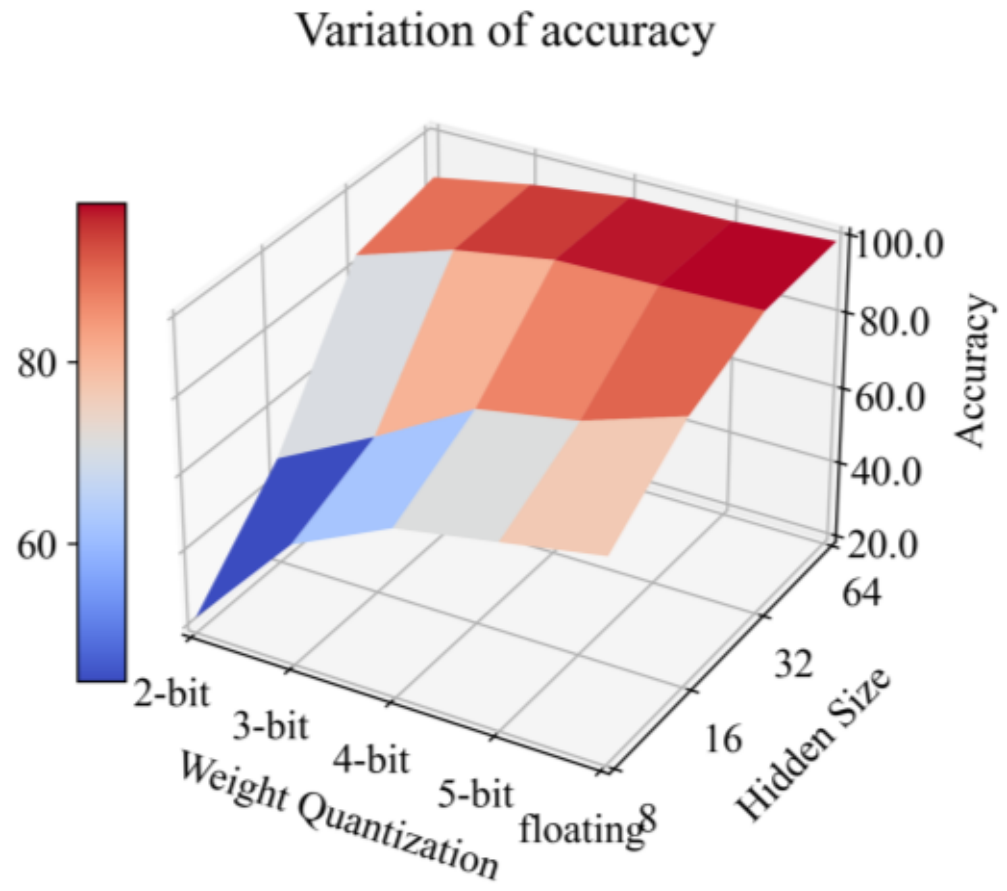

| Hidden size | Parameters |
|---|---|
| 8 | 616 |
| 16 | 2256 |
| 32 | 8608 |
| 64 | 33600 |

**Figure 5.1:** Validation accuracy based on different weight quantization and number of hidden neurons (left). The total number of training parameters based on various hidden size (right).

The DB-LSTM system for ECG forecasting described in this paper enhances the LSTM cell structure to come out high speed training with high accuracy using simple neural network architecture. This model trained with 280 samples and forecast 280 samples, to attain 0.0352 ± 0.0021 RMSE and 98.1 ± 0.15% accuracy. Moreover, fix-point weight is implemented and can achieve similar performance as the floating weight. As a result, online training is reliable for embedding in hardware to build a model based on the individual ECG properties.

## 5.2 Discussion on ECG Classification

A comparison graph shown in Fig. 5.1 which shows the classification accuracy based on hidden neuron number and different weight quantization. According to the figure, it has shown that higher classification accuracy can be obtained by setting larger hidden neuron number and higher weight quantization. With a smaller number of weight bits, the effeteness of accuracy is stronger due to various of hidden size, which can be explained by the total number of training parameters shown on the right-hand side of Fig. 5.1. Lower weight quantization stands for that the choice of weight is limited, thus by involving larger number of parameters, the error due to quantization can be traded off.

The performance of the proposed model is summarized and compared with other state-of-arts in Table VI. [37] is an improved model based on [22], it significantly decreased number of weight parameters by decomposing fully connected layer and CNN layer into multiple steps. However, the drawbacks of this decomposition method are increasing number of activation functions, normalization processes, convolution operations, number of biases, splitting and concatenate operations, which cannot afford parallel computing, and could increase model latency in hardware implementation. Moreover, this model increased the number of CNN layers and maintain window size of 400.Compared to those designs implemented with FP32 weight resolution, the proposed algorithm used one of the smallest network sizes and achieved comparable accuracy. When the weight is quantized to INT4, the accuracy only has minor degradation of 1.6%. If the weight resolution is further reduced to INT3, the accuracy can still be maintained at 91.8%.

TABLE VI BENCHMARK TABLE FOR CLASSIFICATION

| | [21] | [22] | [23] | [24] | [35] | [36] | [37] | This Work |
|---|---|---|---|---|---|---|---|---|
| Year | 2017 | 2019 | 2021 | 2021 | 2021 | 2021 | 2022 | 2022 |
| Dataset | MIT-BIH | MIT-BIH | MIT-BIH | Fantasia | MIT-BIH | PTB Diagnostic | MIT-BIH | MIT-BIH |
| Topology | 1-D CNN | Deep CNN | 1-D CNN | Deep CNN | SVM+MLP | LSTM | Deep CNN | DB-LSTM |
| Number of Layers | 10 | 10 | 10 | 42 | 4 | 3 | 12 | 4 |
| No. of Weight Parameters | 19,750 | 198,037 | 49,216* | >10M | NA Input:32 | NA | 8,157 | 8,608 |
| Window Size | 260 | 400 | 260 | 160 | NA | NA | 400 | 180 |
| No. of Classes | 5 | 5 | 5 | 40 | 5 | 2 | 5 | 5 |
| Validation Accuracy | 93.47% | 98.40% | 98.12% | 99.5% | 79.33% | 84.63% | 99.10% | 91.8%, 96.2% & 97.5% |
| Weight Resolution | FP32 | FP32 | FP32 | FP32 | FP32 | INT1 | FP32 | INT3, INT4 &FP32 |

* Calculated based on information provided in paper.

# Chapter 6

# Conclusion and Future Work

## 6.1 Conclusion

In this report, an ECG forecasting, and classification model is implemented with DB-LSTM, with fixed-point weight. The model structure is designed with pre-processing and input layer, DB-LSTM layer, fully connected layer particularly for classification task and output layer, from which DB-LSTM cell is using dynamic bias where all states are linked to previous or current cell states. Moreover, INT3 down to INT1 weight quantization are implemented on the forecasting model. Also, INT4 and INT3 weight precision are implemented on the classification model to fit edge devices for further purpose, where INT4 attained satisfactory results that contrast with the FP32. In ECG forecasting application, DB-LSTM can perform online training and multi heartbeat prediction, which yields a fast and stable training duration and higher performance. It obtained 98.86% accuracy for normal ECG signal and 0.00026 NMSE with MIT-BIH database. Multi datasets performance had also prove that the proposed innovative model can attain high quality performance. In ECG classification application, DB-LSTM model can perform 97.5% accuracy with five categories with MIT-BIH database. Compared with conventional LSTM model, the DB-LSTM yields shorter window size and higher performance. Moreover, performance of multi models such as CNN and bidirectional LSTM, had also prove that the proposed innovative model can attain high quality performance.

## 6.2 Future Work

To better fit DB-LSTM algorithm into edge devices, the quantization for inputs and piecewise activate functions need to be adjusted. DigiNet, a simulation platform for neural networks on digital hardware has been uploaded to PyPi, which is continued updating. Right now, it supports:

- LSTM and Fully Connected layer in both software and hardware programming,
- various quantization selection for weights (16-, 12- and 8-bit),
- various fraction width in weights (8-, 6- and 4-bit),
- various piece-wise activation functions (sigmoid and tanh),
- pure software simulation for contrast (FP32)

The future of DigiNet is to integrate multiple neural network layers including conventional layer, attention layer altogether. It enables simulation for different hardware quantization requirement on software, which makes the connection and conversion between AI algorithm and edge devices be stronger and simpler.

The next step is to integrate this real-time classification algorithm with hardware implementation to provide ECG anomaly classification for continuous cardiac monitoring. Right now, the conventional LSTM had been implemented on FPGA and waiting for application testing. And the final goal is to apply online training on edge devices for healthcare monitoring. A detailed timeline is shown below:

| | |
|---|---|
| **By end of PhD Year 2** | • Improve DigiNet package.<br>• Implement LSTM algorithm on FPGA and CMOS.<br>• Prepare a tapeout |
| **PhD Year 3** | • Finish DigiNet package to support most NN layers.<br>• Combine sensors and chips to do real time ECG, breath, and uric acid monitoring.<br>• Design online training for complex algorithm |
| **PhD Year 4** | • Tapeout chip that supports online training |

## 6.3 Challenges

The challenges mainly come from:

- The data transmit between sensors, chips and user end (mobile phones, watches)

  The bio-signal are most in analog manner, ADC is the primary choice but cause high energy consumption. Spiking algorithm is a state-of-the-art method that had been proven to replace the function of ADC. [38] had stated the similar performance between spiking model and FFT and yielded high accuracy in speak digit recognition. The material and inner design of sensors will be supported by IME or IMRE, A*STAR. The info communication between chips and user devices could through Bluetooth or NFC, but a specialized and more effective way could be invented.

- Personalize online training algorithm: trade off among memory size, power consumption and latency

  [39] had provided a clear view to perform online training on chips on both training algorithm and hardware design aspects. However, it only implemented online training on basic recurrent neural network, and it cannot afford complex applications. Based on the existing research, online training based on LSTM together with Bio-signal monitoring would be the eventual objective of PhD journey.

# Publications